\documentclass[12pt]{iopart}
\newcommand{\sect}[1]{\setcounter{equation}{0}\section{#1}}
\newcommand{\bfm}[1]{\mbox{\boldmath${#1}$}}

\usepackage{graphicx}
\begin{document}
\title[Generalized Kinetic Equations for a System of Interacting Atoms and Photons]
{Generalized Kinetic Equations for a System of Interacting Atoms
and Photons:\\ Theory and Simulations}
\author{A. Rossani$^\dag$, and A.M. Scarfone$^{\dag,\ddag}$}
\address{\dag\ Istituto Nazionale di Fisica della Materia,
Dipartimento di Fisica - Politecnico di Torino, Corso Duca degli
Abruzzi 24, 10129 Torino, Italy.}
\address{\ddag\ Istituto Nazionale di
Fisica Nucleare, Dipartimento di Fisica - Universit\`a di
Cagliari, Cittadella Universitaria, 09042 Monserrato Cagliari,
Italy.}
\date{\today}
\begin {abstract}
In the present paper we introduce generalized kinetic equations
describing the dynamics of a system of interacting gas and photons
obeying to a very general statistics. In the space homogeneous
case we study the equilibrium state of the system and investigate
its stability by means of Lyapounov's theory. Two physically
relevant situations are discussed in details: photons in a
background gas and atoms in a background radiation. After having
dropped the statistics generalization for atoms but keeping the
statistics generalization for photons, in the zero order
Chapmann-Enskog approximation, we present two numerical
simulations where the system, initially at equilibrium, is
perturbed by an external isotropic Dirac's delta and by a constant
source of photons.
\end {abstract}
\submitto{JPA}
%\keywords{Kinetic theory; Boltzmann equation;
%Gas-Photons interactions; Dynamical systems; Lyapounov theory}
\eads{\mailto{alberto.rossani@polito.it; corresponding
author},\mailto{ antonio.scarfone@polito.it}} \maketitle

%%%%%%%%%%%%%%%%%%%%%%%%%%%%%%%%%%%%%%%%%%%%%%%%%%%%%%%%%%%%%%%%%%%%%%%%%%%%%%%%%%%%%%
\sect{Introduction} In several fields of nuclear and condensed
matter physics, as well as in astrophysics and plasma physics,
there exists experimental evidences which suggest strongly the
necessity of introducing new statistics, different from the
standard
Boltzmann-Gibbs, Bose-Einstein and Fermi-Dirac ones.\\
Typically, generalized statistical distributions arise in systems
exhibiting long-time microscopic memory, long-range interaction or
fractal space-time constraints
\cite{Abe,Kaniadakis0,Kaniadakis00}. Solar neutrinos
\cite{Lavagno1}, high energy nuclear collisions \cite{Lavagno2},
cosmic microwave background radiation \cite{Tsallis1,11} are some
typical phenomena where generalized statistical theories has been
applied. Several models of $q$-deformed statistics have been used
in the study of Bose gas condensation \cite{Hsu,Rego}, phonon
spectrum in $^4$He \cite{Rego1} as well as in the study of
asymmetric $XXZ$ Heisenberg chain \cite{Alcaraz}. Moreover,
quantum $q$-statistics has been introduced in the blackbody theory
\cite{W1,W2} where, based on an equilibrium statistical approach,
a deformed Planck's law has been
derived.\\
Recently, in \cite{RK}, a generalized kinetic theory has been
proposed by Rossani and Kaniadakis, with the purpose to deal, at a
kinetic level, particles obeying to a generalized statistics.
Following this idea, generalized kinetic theories of electrons and
photons \cite{RSc} as well as
electrons and phonons \cite{Ro00} have been proposed.\\
In this paper we present, as a natural continuation of the
previous works, a kinetic theory of interacting atoms
and photons obeying to a very general statistics.\\
We recall that in the last years a classical kinetic approach in
the study of this dynamical problem has been proposed for the case
of two energy levels \cite{RSM,RR} and further generalized for
multi-levels atoms and multi-frequencies photons \cite{RS1,RS2}.
The most remarkable feature is that this approach allows a
self-consistent derivation, at equilibrium, of Planck's law.\\
The physical system we deal with is constituted by atoms
$A_{_\ell}$ ($\ell=1,\,2,\,\cdots,\,N$) with mass $m$, endowed
with a finite number $n$ of internal energy levels
$0=E_{_1}<E_{_2}<\cdots<E_{_n}$, with transitions from one state
to another made possible by either scattering between particles,
or by their interaction with a self-consistent radiation field
made up by photons $p_{_{ij}}$ with intensity $I_{_{ij}}$,
$i<j=1,\,2,\,\cdots,\,n$, at the $n\,(n-1)/2$ frequencies
$\omega_{_{ij}}=E_{_j}-E_{_i}$
(we adopt natural units $\hbar=c=1$).\\
According to the relevant literature \cite{Ox}, we assume that
the following interactions take place:\\
a) elastic and inelastic interactions between particles,\\
b) emission and absorption of photons.\\
We restrict ourselves to the most common and physical interaction
mechanism between particles:
\begin{eqnarray}
A_{_k}+A_{_i}\rightleftharpoons A_{_k}+A_{_j} \ ,\label{re}
\end{eqnarray}
where $i\leq j$. The most general reaction \cite{RS1} does not
introduce significant improvement with respect to the present
simplification.\\ Gas-radiation interaction processes include:
\begin{enumerate}
\item absorption:
\begin{eqnarray}
A_{_i}+p_{_{ij}}\rightarrow A_{_j} \ ;
\end{eqnarray}
\item spontaneous emission:
\begin{eqnarray}
A_{_j}\rightarrow A_{_i}+p_{_{ij}} \ ;
\end{eqnarray}
\item stimulated emission:
\begin{eqnarray}
A_{_j}+p_{_{ij}}\rightarrow A_{_i}+2\,p_{_{ij}} \ .
\end{eqnarray}
\end{enumerate}
Following Ref.s \cite{RSM,RS1}, for these emission and absorption
processes we assume that: the outcoming atom has the same
velocity as the incoming atom (that is we neglect photon momentum
with respect to atom momentum); photons are spontaneously emitted
isotropically; photons which derive from stimulated emission
have the same directions as the incoming photons.\\
Finally, gas-radiation interactions are modeled by means of the
Einstein coefficients \cite{Po}: $\beta_{_{ij}}$ for both
absorption and stimulated emission, $\alpha_{_{ij}}$ for
spontaneous emission.\\

The paper is so organized. In the next Section 2, we introduce the
kinetic equation for the physical system of interacting atoms and
photons. In Sections 3 and 4 theoretical results on equilibrium of
the deformed model and its stability, by means of Lyapounov's
theory, are given. The mathematical results are discussed on a
physical ground, and connections with thermodynamics are pointed
out. Two particular situations are studied in Section 5 and 6:
photons in a background of atoms and atoms in a background
radiation. In Section 7, we drop the statistics generalization for
atoms but keep the generalization for photons. The results of two
numerical simulations are presented. In the first simulation, the
system initially in equilibrium, is disturbed, starting from time
$t=0$, by a flash of monochromatic photons injected by an
isotropic external source. In the second simulation, an isotropic
source of monochromatic photons, constant in the time, is inserted
starting from time $t=0$. Section 8 deals with the conclusions.
Finally, in the Appendix A we present an overall sum up on the
kinetic equations for a system of atoms and photons obeying to the
standard statistics while, in the Appendix B we recall briefly two
generalized statistical distributions used in the numerical
simulations.

%%%%%%%%%%%%%%%%%%%%%%%%%%%%%%%%%%%%%%%%%%%%%%%%%%%%%%%%%%%%%%%%%%%%%%%%%%%%%%%%%%%%%%
\sect{Generalized Boltzmann equations} In this section we
introduce the kinetic equation describing a system of atoms and
photons obeying to a very general statistics. For sake of clarity
we begin by introducing the \"{U}hling-Uhlenbeck
equation (UUe) in order to fix the notations used in the following.\\
The UUe is a quantum extension of the Boltzmann equation
\cite{Ue}. Let us consider a system of particles with mass $m$
which obeies the Bose-Einstein or Fermi-Dirac statistics. These
particles interact elastically by means of binary encounters. In
this situation the UUe reads as
\begin{eqnarray}
\nonumber \fl\hspace{15mm}\frac{\partial\,f}{\partial\,t}+{\bfm
v}\cdot{\bfm\nabla}f&=&\int_{I\!\!R^3\times
S^2}g\,\sigma(g,\,\zeta)\Bigg\{f({\bfm v}^\prime)\,f({\bfm
w}^\prime)
\,\Big[1+\lambda\,f({\bfm v})\Big]\,\Big[1+\lambda\,f({\bfm w})\Big]\\
&-& f({\bfm v})\,f({\bfm w})\,\Big[1+\lambda\,f({\bfm
v}^\prime)\Big]\,\Big[1+\lambda\,f({\bfm
w^\prime})\Big]\Bigg\}\,d{\bfm w}\,d{\bfm\Omega}^\prime \
,\label{UU}
\end{eqnarray}
where $\lambda=1$ for bosons and $\lambda=-1$ for fermions. In Eq.
(\ref{UU}) $f({\bfm v})\equiv f({\bfm x},\,{\bfm v},\,t)$ is the
dimensionless distribution function, $\sigma(g,\,\zeta)$ is the
cross section, with $g=|{\bfm v}-{\bfm w}|$ the relative speed.
The post-collision velocities are given by
\begin{eqnarray}
{\bfm v}^\prime={1\over2}\,({\bfm v}+{\bfm
w}+g\,{\bfm\Omega}^\prime) \ ,\hspace{10mm}{\bfm
w}^\prime={1\over2}\,({\bfm v}+{\bfm w}-g\,{\bfm\Omega}^\prime) \
,
\end{eqnarray}
where $\bfm v$ and $\bfm w$ are the velocities of the incoming
particles and $\cos\zeta={\bfm\Omega}\cdot{\bfm\Omega}^\prime$,
with ${\bfm\Omega}=({\bfm v}-{\bfm w})/g$ the unit vector in the
direction of the relative speed. Finally the two-dimensional unit
sphere $S^2$ is the domain of integration
for the unit vector ${\bfm\Omega}^\prime$.\\
In view of applications to particles which obey a more general
statistics, we introduce the following substitution in the
collision integral
\begin{equation}
f({\bfm v})\rightarrow\varphi[f({\bfm v})] \
,\hspace{20mm}1+\lambda\,f({\bfm v})\rightarrow\psi[f({\bfm v})] \
,
\end{equation}
where $\varphi(x)$ and $\psi(x)$ are non-negative functions
supposed to be continuous as their derivatives. The limit values
for $x\rightarrow0$ are subject to precise physical requirement.
For instance, $\varphi(0)=0$ means that no transition occurs when
one of the initial states is empty, $\psi(0)=1$ means that no
inhibition occurs when one of the final states is empty. According
to \cite{Ro00}, we assume
\begin{eqnarray}
\frac{d}{d\,x}\left({\varphi(x)\over\psi(x)}\right)>0 \
.\label{con}
\end{eqnarray}
This is trivially true for bosons and fermions. In general this
assumption is justified a posteriori since it assures uniqueness
and stability of equilibrium.\\
The generalized UUe now reads as follows
\begin{eqnarray}
\nonumber \fl\hspace{15mm}\frac{\partial\,f}{\partial\,t}+{\bfm
v}\cdot{\bfm\nabla}f&=&\int_{I\!\!R^3\times
S^2}g\,\sigma(g,\,\zeta)\Bigg\{\varphi[f({\bfm
v}^\prime)]\,\varphi[f({\bfm w}^\prime)]
\,\psi[f({\bfm v})]\,\psi[f({\bfm w})]\\
&-& \varphi[f({\bfm v})]\,\varphi[f({\bfm w})]\,\psi[f({\bfm
v}^\prime)]\,\psi[f({\bfm w^\prime})]\Bigg\}\,d{\bfm
w}\,d{\bfm\Omega}^\prime \ .\label{gUU}
\end{eqnarray}

In the following we specify Eq. (\ref{gUU}) for a system of
interacting atoms and photons. Let us introduce the characteristic
departure and arrival functions \cite{RK}: $\varphi(f_{_\ell})$
and $\psi(f_{_\ell})$ for atoms, as well as $\Phi(I_{_{ij}})$ and
$\Psi(I_{_{ij}})$ for photons, where atoms $A_{_\ell}$, endowed
with energy level $E_{_\ell}$, are described by means of the
distribution function $f_{_\ell}({\bfm x},\,{\bfm v},\,t)$, and
photons $p_{_{ij}}$, at frequency $\omega_{_{ij}}$, are described
by means of the radiation intensity $I_{_{ij}}({\bfm
x},\,{\bfm\Omega},\,t)$.\\  The distribution function of atoms
$A_\ell$ obeys to the following system of generalized Boltzmann
equations (see Appendix A) \cite{RS1}
\begin{eqnarray}
\frac{\partial\,f_{_\ell}}{\partial\,t}+{\bfm
v}\cdot{\bfm\nabla}f_{_\ell}=J_{_\ell}({\bfm v})+{\widetilde
J}_{_\ell}({\bfm v}) \ ,\label{boltzmann}
\end{eqnarray}
where in the right hand side of Eq. (\ref{boltzmann})
$J_{_\ell}({\bfm v})$ describes the contribution due to the
atom-atom interaction and ${\widetilde J}_{_\ell}({\bfm v})$ the
contribution due to the atom-photon
interaction.\\
Explicitly, the collision integral, which takes into account the
inelastic and elastic atom-atom collision, is given by
\begin{eqnarray}
\fl\hspace{8mm}J_{_\ell}({\bfm v})=\sum_j\sum_{i\leq
j}J^{^{(1)}}_{_{ij\ell}}({\bfm v})+\sum_j\sum_{i\leq
j}J^{^{(2)}}_{_{ij\ell}}({\bfm v})+\sum_k\sum_{j\geq
\ell}J^{^{(3)}}_{_{jk\ell}}({\bfm v})+\sum_k\sum_{i\leq
\ell}J^{^{(4)}}_{_{ik\ell}}({\bfm v}) \ ,\label{atoms}
\end{eqnarray}
with
\begin{eqnarray}
\nonumber J^{^{(1)}}_{_{ij\ell}}({\bfm v})&=&\int_{I\!\!R^3\times
S^2} g\,\sigma_{_{\ell j}}^{^{\ell i}}(g,\,\zeta)\,
\Bigg\{\varphi[f_{_\ell}({\bfm
v}^+_{_{ij}})]\,\varphi[f_{_i}({\bfm
w}^+_{_{ij}})]\,\psi[f_{_\ell}({\bfm v})]\,\psi[f_{_j}({\bfm
w})]\\
&-&\varphi[f_{_\ell}({\bfm v})]\,\varphi[f_{_j}({\bfm
w})]\,\psi[f_{_\ell}({\bfm v_{_{ij}}^+})]\,\psi[f_{_i}({\bfm
w_{_{ij}}^+})]\Bigg\}\,d{\bfm w}\,d{\bfm\Omega}^\prime \ ,
\end{eqnarray}
\begin{eqnarray}
\nonumber J^{^{(2)}}_{_{ij\ell}}({\bfm v})&=&\int_{I\!\!R^3\times
S^2} g\,\sigma_{_{\ell i}}^{^{\ell
j}}(g,\,\zeta)\Bigg\{\varphi[f_{_\ell}({\bfm
v}^-_{_{ij}})]\,\varphi[f_{_j}({\bfm
w}^-_{_{ij}})]\,\psi[f_{_\ell}({\bfm v})]\,\psi[f_{_i}({\bfm
w})]\\&-&\varphi[f_{_\ell}({\bfm v})]\,\varphi[f_{_i}({\bfm
w})]\,\psi[f_{_\ell}({\bfm v_{_{ij}}^-})]\,\psi[f_{_j}({\bfm
w_{_{ij}}^-})]\Bigg\}\,d{\bfm w}\,d{\bfm\Omega}^\prime \
,\label{j2}
\end{eqnarray}
\begin{eqnarray} \nonumber
J^{^{(3)}}_{_{jk\ell}}({\bfm v})&=&\int_{I\!\!R^3\times S^2}
g\,\sigma_{_{k\ell
}}^{^{kj}}(g,\,\zeta)\Bigg\{\varphi[f_{_j}({\bfm v}^-_{_{\ell
j}})]\,\varphi[f_{_k}({\bfm w}^-_{_{\ell
j}})]\,\psi[f_{_\ell}({\bfm v})]\,\psi[f_{_k}({\bfm
w})]\\&-&\varphi[f_{_\ell}({\bfm v})]\,\varphi[f_{_k}({\bfm
w})]\,\psi[f_{_j}({\bfm v_{_{\ell j}}^-})]\,\psi[f_{_k}({\bfm
w_{_{\ell j}}^-})]\Bigg\}\,d{\bfm w}\,d{\bfm\Omega}^\prime \
,\label{j3}
\end{eqnarray}
\begin{eqnarray}
\nonumber J^{^{(4)}}_{_{ik\ell}}({\bfm v})&=&\int_{I\!\!R^3\times
S^2}
g\,\sigma_{_{k\ell}}^{^{ki}}(g,\,\zeta)\Bigg\{\varphi[f_{_i}({\bfm
v}^+_{_{i\ell}})]\,\varphi[f_{_k}({\bfm
w}^+_{_{i\ell}})]\,\psi[f_{_\ell}({\bfm v})]\,\psi[f_{_k}({\bfm
w})]\\&-&\varphi[f_{_\ell}({\bfm v})]\,\varphi[f_{_k}({\bfm
w})]\,\psi[f_{_i}({\bfm v_{_{i\ell}}^+})]\,\psi[f_{_k}({\bfm
w_{_{i\ell}}^+})]\Bigg\}\,d{\bfm w}\,d{\bfm\Omega}^\prime \ ,\
\end{eqnarray}
where $\sigma_{_{\ell i}}^{^{\ell j}}$ and $\sigma_{_{\ell
j}}^{^{\ell i}}$ are the cross sections for forward and backward
reactions describing elastic and inelastic interactions. They
satisfy the following microreversibility relationships \cite{RS1}
\begin{eqnarray}
\fl\hspace{10mm}g^2\sigma_{_{\ell i}}^{^{\ell
j}}(g,\,{\bfm\Omega}^\prime)=(g_{_{ij}}^-)^2\,\sigma_{_{\ell
j}}^{^{\ell i}}(g_{_{ij}}^-,\,{\bfm\Omega}^\prime) \
,\hspace{10mm} g^2\,\sigma_{_{\ell j}}^{^{\ell
i}}(g,\,{\bfm\Omega}^\prime)=(g_{_{ij}}^+)^2\,\sigma_{_{\ell
i}}^{^{\ell j}}(g_{_{ij}}^+,\,{\bfm\Omega}^\prime) \
.\label{micro}
\end{eqnarray}
The post-collision velocities are defined by
\begin{eqnarray}
{\bfm v}^\pm_{_{ij}}={1\over2}\,({\bfm v}+{\bfm
w}+g_{_{ij}}^\pm\,{\bfm\Omega}^\prime) \ ,\hspace{10mm}{\bfm
w}^\pm_{_{ij}}={1\over2}\,({\bfm v}+{\bfm
w}-g_{_{ij}}^\pm\,{\bfm\Omega}^\prime) \ ,\label{v}
\end{eqnarray}
with
\begin{eqnarray}
g_{_{ij}}^\pm=\sqrt{g^2\pm{4\over m}\,(E_{_j}-E_{_i})} \ .
\end{eqnarray}
The four contributions to $J_{_\ell}({\bfm v})$ correspond to the
cases in which $A_{_\ell}$ plays the role, in the reaction scheme
(\ref{re}), of $A_{_k}$ in the r.h.s., $A_{_k}$ in the l.h.s.,
$A_{_i}$ and $A_{_j}$, respectively.

Differently, the collision integral ${\widetilde J}_{_\ell}({\bfm
v})$, which accounts for the gas-radiation interactions, is given
by
\begin{eqnarray}
{\widetilde J}_{_\ell}({\bfm v})=\sum_{i>\ell}\int_{S^2}{\widehat
J}_{_{\ell i}}({\bfm
v},\,{\bfm\Omega})\,d{\bfm\Omega}-\sum_{i<\ell}\int_{S^2}{\widehat
J}_{_{i\ell}}({\bfm v},\,{\bfm\Omega})\,d{\bfm\Omega} \
,\label{radiation}
\end{eqnarray}
where
\begin{eqnarray}
\fl\hspace{10mm}{\widehat J}_{_{i\ell}}({\bfm
v},\,{\bfm\Omega})=\alpha_{_{i\ell
}}\,\Bigg\{\varphi[f_{_\ell}({\bfm v})]\,\Psi[I_{_{i\ell
}}({\bfm\Omega})]\,\psi[f_{_i}({\bfm v})]-\varphi[f_{_i}({\bfm
v})]\,\Phi[I_{_{i\ell}}({\bfm\Omega})]\,\psi[f_{_\ell}({\bfm
v})]\Bigg\} \ .\label{rad}
\end{eqnarray}
By taking into account all the energy levels higher than $\ell$,
the loss term is due to absorption, while the gain term is due to
spontaneous and stimulated emission. The situation is reversed
when we consider all the energy levels lower than $\ell$.\\

Finally, the kinetic equation for photons $p_{ij}$ reads:
\begin{eqnarray}
\frac{\partial\,I_{_{ij}}}{\partial\,t}+{\bfm\Omega}\cdot{\bfm\nabla}I_{_{ij}}=
\omega_{_{ij}}\,{\widetilde J}_{_{ij}}({\bfm\Omega}) \ ,
\label{kinetic}
\end{eqnarray}
where
\begin{eqnarray}
{\widetilde J}_{_{ij}}({\bfm\Omega})=\int_{I\!\!R^3}{\widehat
J}_{_{ij}}({\bfm v},\,{\bfm\Omega})\,d{\bfm v} \ .\label{rad1}
\end{eqnarray}
Again, the gain term is due to spontaneous and stimulated
emission, while the loss term is due to absorption. \\
It is easy to see that, by posing
\begin{eqnarray}
&&\varphi[f_{_\ell}[({\bfm v})]\rightarrow f_{_\ell}({\bfm v}) ,
\hspace{20mm}
\psi[f_{_\ell}({\bfm v})]\rightarrow1 \ ,\label{clas}\\
&&\Phi[I_{_{ij}}({\bfm\Omega})]\rightarrow
\frac{\beta_{_{ij}}}{\alpha_{_{ij}}}\,I_{_{ij}}({\bfm\Omega}) ,
\hspace{11mm} \Psi[I_{_{ij}}({\bfm\Omega})]\rightarrow
1+\frac{\beta_{_{ij}}}{\alpha_{_{ij}}}\,I_{_{ij}}({\bfm\Omega}) \
,\label{clas1}
\end{eqnarray}
Eq.s (\ref{boltzmann}) and (\ref{kinetic}) reduce to the standard
kinetic equations for atoms and photons \cite{RS1,RS2}.\\ From now
on, for sake of simplicity, we adopt the notations
$\varphi_{_\ell}({\bfm v})\equiv\varphi[f_{_\ell}({\bfm v})]$,
$\psi_{_\ell}({\bfm v})\equiv\psi[f_{_\ell}({\bfm v})]$,
$\Phi_{_{ij}}({\bfm \Omega})\equiv\Phi[I_{_{ij}}({\bfm\Omega})]$
and
$\Psi_{_{ij}}({\bfm\Omega})\equiv\Psi[I_{_{ij}}({\bfm\Omega})]$.
%%%%%%%%%%%%%%%%%%%%%%%%%%%%%%%%%%%%%%%%%%%%%%%%%%%%%%%%%%%%%%%%%%%%%%%%%%%%%%%%%%%%%%
\sect{Equilibrium} Notwithstanding the generalized equations are
more complicate with respect to the classical kinetic equations,
many of the methods used in the standard kinetic theory are still
applicable. This is in particular true for the study of
equilibria and their stability.\\

 {\bf Lemma 1.} For any given
arbitrary smooth functions $\gamma_{_\ell}({\bfm v})$ and
$\Gamma_{_{ij}}({\bfm \Omega})$ the following two relations hold
\begin{eqnarray}
\nonumber \fl\sum_\ell\int_{I\!\!R^3}\gamma_{_\ell}({\bfm v})
\,J_{_{\ell}}({\bfm v}) \,d{\bfm v}=\sum_\ell\sum_j\sum
_{i\le j}\int_{I\!\!R^3\times I\!\!R^3\times S^2}g\,\sigma^{^{\ell
i}}_{_{\ell j}}(g,\,\zeta)\\
\nonumber\times\left[\varphi_{_\ell}({\bfm v}_{_{ij}}^+)\,
\varphi_{_i}({\bfm w}_{_{ij}}^+)\,\psi_{_\ell}({\bfm
v})\,\psi_{_j}({\bfm w})-\varphi_{_\ell}({\bfm
v})\,\varphi_{_j}({\bfm w})\,\psi_{_\ell}({\bfm
v}_{_{ij}}^+)\,\psi_{_i}({\bfm
w}_{_{ij}}^+)\right]\\
\times\left[\gamma_{_\ell}({\bfm v})+\gamma_{_j}({\bfm
w})-\gamma_{_\ell}({\bfm v}_{_{ij}}^+)-\gamma_{_i}({\bfm
w}_{_{ij}}^+)\right]\,d{\bfm v}\,d{\bfm w}\,d{\bfm\Omega}^\prime
\ ,\label{l1}
\end{eqnarray}
\begin{eqnarray}
\nonumber \fl\sum_\ell\int_{I\!\!R^3}\gamma_{_\ell}({\bfm
v})\,{\widetilde J}_{_\ell}({\bfm v})\,d{\bfm
v}+\sum_j\sum_{i<j}\int_{
S^2}\Gamma_{_{ij}}({\bfm\Omega})\,{\widetilde J}_{_{ij}}({\bfm
\Omega})\,d{\bfm\Omega}\\
\nonumber =\sum_j\sum_{i<j}\alpha_{_{ij}}\int_{I\!\!R^3\times
S^2}\left[\varphi_{_j}({\bfm v})\,\Psi_{_{ij}}({\bfm \Omega}
)\,\psi_{_i}({\bfm v})-\varphi_{_i}({\bfm
v})\,\Phi_{_{ij}}({\bfm\Omega})\,\psi_{_j}({\bfm v})\right]\\
\times\left[ \gamma_{_i}({\bfm
v})+\Gamma_{_{ij}}({\bfm\Omega})-\gamma_{_j}({\bfm
v})\right]\,d{\bfm v}\,d{\bfm\Omega} \ .\label{l2}
\end{eqnarray}
\\

{\it Proof.} The proof for the generalized case follows the same
steps which can be found in Ref. \cite{RS1} for the standard
theory. In the following we present the main points recalling to
the relevant references for the details \cite{RS1,Liboff,Chapman}.\\
Eq. (\ref{l1}) follows from the definition of the collision
integral $J_{_{\ell}}({\bfm v})$ given in Eq. (\ref{atoms}). The
relevant addends may be grouped together, with suitable interchang
of the indices $i$, $j$, $k$ and $\ell$, to form the collision
term appearing on the r.h.s of Eq. (\ref{l1}). In particular, for
all the $J^{^{(2)}}_{_{ij\ell}}({\bfm v})$ and
$J^{^{(3)}}_{_{jk\ell}}({\bfm v})$ the microreversibility
relationships (\ref{micro}) are used as well as the relation
\begin{eqnarray}
\frac{g_{_{ij}}^\pm}{g}\,d{\bfm v}\,d{\bfm
w}\,d{\bfm\Omega}^\prime=d{\bfm v_{_{ij}}^\pm}\,d{\bfm
w_{_{ij}}^\pm}\,d{\bfm\Omega} \ ,\label{jac}
\end{eqnarray}
where the Jacobian of the transformation in Eq. (\ref{jac}) arises
from the definitions (\ref{v}).\\ In the same way, Eq. (\ref{l2})
is obtained from the collision integrals ${\widetilde
J}_{_\ell}({\bfm v})$ and ${\widetilde J}_{_{ij}}({\bfm \Omega})$
given in Eq.s (\ref{radiation}) and (\ref{rad1}), after suitable
interchange of the indices $i$, $j$ and
$\ell$. $\ \bullet$\\

In the space homogeneous case, where all the dynamical quantities
$f_{_\ell}$ and $I_{_{ij}}$ are functions only of the time,
equilibrium is defined by
\begin{eqnarray}
{\partial\,f_{_\ell}\over\partial\, t}={\partial\,
I_{_{ij}}\over\partial\, t}=0 \ .\label{equi}
\end{eqnarray}
Let us introduce the following functional:
\begin{eqnarray}
\fl\hspace{15mm}{\cal
D}=\sum_\ell\int_{I\!\!R^3}\ln\left({\varphi_{_\ell}\over\psi_{_\ell}}\right)
\frac{\partial\,f_{_\ell}}{\partial\, t}\,\,d{\bfm
v}+\sum_j\sum_{i<j}{1\over \omega_{_{ij}}}\int_{ S^2}
\ln\left({\Phi_{_{ij}}\over\Psi_{_{ij}}}\right)\,{\partial
\,I_{_{ij}}\over\partial\,t}\,d{\bfm\Omega} \ .\label{dp}
\end{eqnarray}
After using the kinetic equations (\ref{boltzmann}) and
(\ref{kinetic}), Eq. (\ref{dp}) becomes
\begin{eqnarray}
\nonumber\fl\hspace{15mm}{\cal
D}=\sum_\ell\int_{I\!\!R^3}\ln\left({\varphi_{_\ell}\over\psi_{_\ell}}\right)
\,J_{_\ell}({\bfm v})\,d{\bfm
v}\\+\sum_\ell\int_{I\!\!R^3}\ln\left({\varphi_{_\ell}\over\psi_{_\ell}}\right)
\,{\widetilde J}_{_\ell}({\bfm v})\,d{\bfm
v}+\sum_j\sum_{i<j}\int_{ S^2}
\ln\left({\Phi_{_{ij}}\over\Psi_{_{ij}}}\right)\,{\widetilde
J}_{_{ij}}({\bfm\Omega})\,d{\bfm\Omega} \ ,
\end{eqnarray}
and by applying Lemma 1, with
$\gamma_{_\ell}=\ln(\varphi_{_\ell}/\psi_{_\ell})$ and
$\Gamma_{_{ij}}=\ln(\Phi_{_{ij}}/\Psi_{_{ij}})$, we obtain
\begin{eqnarray}
\nonumber \fl\hspace{15mm} {\cal D}=\sum_\ell\sum_j\sum_{i\leq j
}\int_{I\!\!R^3\times I\!\!R^3\times S^2}
g\,\sigma_{_{\ell j}}^{^{\ell i}}(g,\,\zeta)\\
\nonumber\fl\hspace{15mm}\times\left[\varphi_{_\ell}({\bfm
v}_{_{ij}}^+)\, \varphi_{_i}({\bfm
w}_{_{ij}}^+)\,\psi_{_\ell}({\bfm v})\,\psi_{_j}({\bfm
w})-\varphi_{_\ell}({\bfm v})\,\varphi_{_j}({\bfm
w})\,\psi_{_\ell}({\bfm
v}_{_{ij}}^+)\,\psi_{_i}({\bfm w}_{_{ij}}^+)\right]\\
\nonumber\fl\hspace{15mm}\times\ln{\varphi_{_\ell}({\bfm
v})\,\varphi_{_j}({\bfm w})\,\psi_{_\ell}({\bfm
v}_{_{ij}}^+)\,\psi_{_i}({\bfm w}_{ij}^+)\over
\varphi_{_\ell}({\bfm v}_{_{ij}}^+)\, \varphi_{_i}({\bfm
w}_{_{ij}}^+)\,\psi_{_\ell}({\bfm
v})\,\psi_{_j}({\bfm w})}\,d{\bfm v}\,d{\bfm w}\,d{\bfm\Omega}^\prime\\
\nonumber\fl\hspace{15mm}+\sum_j\sum_{i<j}\alpha_{_{ij}}\int_{I\!\!R^3\times
 S^2}\,\left[\varphi_{_j}({\bfm
v})\,\Psi_{_{ij}}({\bfm \Omega} )\,\psi_{_i}({\bfm
v})-\varphi_{_i}({\bfm
v})\,\Phi_{_{ij}}({\bfm\Omega})\,\psi_{_j}({\bfm v})\right]\\
\fl\hspace{15mm}\times\ln{\varphi_{_i}({\bfm
v})\,\Phi_{_{ij}}({\bfm\Omega})\,\psi_{_j}({\bfm v})\over
\varphi_{_j}({\bfm
v})\,\Psi_{_{ij}}({\bfm\Omega})\,\psi_{_i}({\bfm v})}\,d{\bfm
v}\,d{\bfm\Omega} \ ,\label{d}
\end{eqnarray}
which is a quantity
manifestly not positive since $(A-B)\,\ln(B/A)\leq 0$.\\

{\bf Proposition 1.} Condition (\ref{equi}) is equivalent to the
following couple of equations
\begin{eqnarray}
\varphi_{_\ell}({\bfm v}_{_{ij}}^+)\,\varphi_{_i}({\bfm
w}_{_{ij}}^+)\,\psi_{_\ell}({\bfm v})\,\psi_{_j}({\bfm
w})=\varphi_{_\ell}({\bfm v})\,\varphi_{_j}({\bfm
w})\,\psi_{_\ell}({\bfm v}_{_{ij}}^+)\,\psi_{_i}({\bfm
w}_{_{ij}}^+) \ ,\label{c1}\\
\hspace{16mm}\varphi_{_j}({\bfm
v})\,\Psi_{_{ij}}({\bfm\Omega})\,\psi_{_i}({\bfm
v})=\varphi_{_i}({\bfm
v})\,\Phi_{_{ij}}({\bfm\Omega})\,\psi_{_j}({\bfm v}) \ .\label{c2}
\end{eqnarray}
\\
{\it Proof}. Taking into account the definitions of the collision
integral given in Eq.s (\ref{atoms}), (\ref{radiation}) and
(\ref{rad1}), we observe first that Eq.s (\ref{c1}) and (\ref{c2})
imply $J_{_\ell}({\bfm v})=\widetilde J_{_\ell}({\bfm
v})=\widetilde J_{_{ij}}({\bfm \Omega})=0$ and thus, from the
kinetic equations (\ref{boltzmann}) and (\ref{kinetic}), we obtain
condition (\ref{equi}).\\ On the other hand, from Eq. (\ref{dp})
by using Eq. (\ref{equi}), it follows that ${\cal D}=0$. But,
since in Eq. (\ref{d}) both the integrands are never positive,
condition ${\cal D}=0$ implies Eq.s (\ref{c1}) and
(\ref{c2}). $\bullet$\\

In the case $i=j$, Eq. (\ref{c1}) gives
\begin{eqnarray}
\ln\left(\frac{\varphi_{_\ell}({\bfm v})}{\psi_{_\ell}({\bfm
v})}\right)+\ln\left(\frac{\varphi_{_i}({\bfm w})}{\psi_{_i}({\bfm
w})}\right)= \ln\left(\frac{\varphi_{_\ell}({\bfm
v}_{_{ii}}^+)}{\psi_{_\ell}({\bfm
v}_{_{ii}}^+)}\right)+\ln\left(\frac{\varphi_{_i}({\bfm
w}_{_{ii}}^+)}{\psi_{_i}({\bfm w}_{_{ii}}^+)}\right) \ ,
\end{eqnarray}
therefore $\ln(\varphi_{_\ell}/\psi_{_\ell})$ is a collision
invariant for atoms, that is
\begin{eqnarray}
\ln\left({\varphi_{_i}\over\psi_{_i}}\right)=-{1\over
T}\,\left({1\over2}\,m\,{\bfm v}^2+E_{_i}-\mu_{_i}\right) \
,\label{a}
\end{eqnarray}
where $T$ is the absolute temperature of the whole system of atoms
and photons. (Here and in the following we pose $k_{\rm B}=1$).\\
For the case $i\not=j$, from Eq. (\ref{c1}) and by using Eq.
(\ref{a}), we get, as additional condition, that all the chemical
potentials are equal: $\mu_{_i}=\mu_{_j}\equiv\mu$.

Differently, from Eq. (\ref{c2}), using Eq. (\ref{a}), follows

\begin{eqnarray}
\ln\left({\Phi_{_{ij}}\over\Psi_{_{ij}}}\right)=-{\omega_{_{ij}}\over
T} \ ,\label{b}
\end{eqnarray}
which is the generalized version of Planck's law. In Appendix B we
write explicitly Eq. (\ref{b}) for two particular statistics.

Let us to observe that, due to the monotonicity of both
$\varphi_{_\ell}/\psi_{_\ell}$ and $\Phi_{_{ij}}/\Psi_{_{ij}}$,
Eq.s (\ref{a}) and (\ref{b}) give an unique equilibrium solutions
for $f_{_\ell}({\bfm v})$ and $I_{_{ij}}({\bfm\Omega})$,
respectively.
%%%%%%%%%%%%%%%%%%%%%%%%%%%%%%%%%%%%%%%%%%%%%%%%%%%%%%%%%%%%%%%%%%%%%%%%%%%%%%%%%%%%%%
\sect{Stability}

In order to study the stability of such equilibrium solution let
us introduce the following functional
\begin{eqnarray}
L=H_{\rm A}+H_{\rm p} \ ,\label{L}
\end{eqnarray}
where
\begin{eqnarray}
H_{\rm A}=\sum_\ell\int_{I\!\!R^3}{\cal H}_{\rm
A}(f_{_\ell})\,d{\bfm v} \
,\label{ha}\\
H_{\rm p}=\sum_j\sum_{i<j}{1\over \omega_{_{ij}} }\int_{S^2}{\cal
H}_{\rm p}(I_{_{ij}})\,d{\bfm\Omega} \ ,\label{hp}
\end{eqnarray}
with
\begin{eqnarray}
{\partial\,{\cal H}_{\rm A}(f_{_\ell})\over\partial\,f_{_\ell}}=
\ln\left({\varphi_{_\ell}\over\psi_{_\ell}}\right) \ ,\label{ha1}\\
{\partial\,{\cal H}_{\rm p}(I_{_{ij}})\over\partial\,I_{_{ij}}}=
\ln\left({\Phi_{_{ij}}\over\Psi_{_{ij}}}\right) \ .\label{hp1}
\end{eqnarray}
Remark that, since $\varphi_{_\ell}/\psi_{_\ell}$ and
$\Phi_{_{ij}}/\Psi_{_{ij}}$ have been assumed to be monotonically
increasing, ${\cal H}_{\rm A}(f_{_\ell})$ and ${\cal H}_{\rm
p}(I_{_{ij}})$ are convex functions
of their arguments.\\
From \cite{RK} we know that $S=-L$ is nothing else but entropy
density for the present physical system and it is the sum of two
contributions: gas entropy and photon entropy
 \cite{Ox,Liboff,Ye}.\\

{\bf Lemma 2.} The condition
\begin{eqnarray}
\nonumber \fl\sum_\ell\int_{I\!\!R^3}\left({\partial\,{\cal
H}_{\rm A}(f_{_\ell})\over\partial\,f_{_\ell}}\right)^*
(f_{_\ell}-f_{_\ell}^*)\,d{\bfm v}+\sum_j\sum_{i<j}{1\over
\omega_{_{ij}}} \int_{S^2}\left({\partial\,{\cal H}_{\rm
p}(I_{_{ij}})\over\partial\,I_{_{ij}}}
\right)^*(I_{_{ij}}-I_{_{ij}}^*)\,d{\bfm\Omega}=0 \ ,\\ \label{h0}
\end{eqnarray}
where $\ast$ means at equilibrium, is equivalent to energy
conservation
$E(f_{_\ell},\,I_{_{ij}})=E^*(f_{_\ell}^\ast,\,I_{_{ij}}^\ast)$.\\

{\it Proof.} Energy can be written as
\begin{eqnarray}
E=E_{\rm A}+E_{\rm p} \ ,
\end{eqnarray}
where
\begin{eqnarray}
E_{\rm A}=\int_{I\!\!R^3}{\cal E}_{\rm A}({\bfm v})\,d{\bfm v} \
,\label{ea}\\
E_{\rm p}=\int_{S^2}{\cal E}_{\rm p}({\bfm\Omega})\,d{\bfm\Omega}
\ ,\label{ep}
\end{eqnarray}
with
\begin{eqnarray}
{\cal E}_{\rm A}=\sum_\ell\left(\frac{1}{2}\,m\,{\bfm v}^2+E_{_\ell}\right)\,f_{_\ell}({\bfm v}) \ ,\\
{\cal E}_{\rm p}= \sum_j\sum_{i<j}I_{_{ij}}\,({\bfm\Omega}) \ .
\end{eqnarray}
By means of Euler's theorem we can write
\begin{eqnarray}
\fl E-E^*=\sum_\ell\int_{I\!\!R^3}\left({\partial\,{\cal E}_{\rm
A}\over\partial\,f_{_\ell}}\right)^*
(f_{_\ell}-f_{_\ell}^*)\,d{\bfm v}
+\sum_j\sum_{i<j}\int_{S^2}\left({\partial\,{\cal E}_{\rm
p}\over\partial\,
I_{_{ij}}}\right)^*\,\left(I_{_{ij}}-I_{_{ij}}^*\right)\,d{\bfm\Omega}
\ .\label{e}
\end{eqnarray}
Differently, by using Eq.s (\ref{a}) and (\ref{b}), from Eq.s
(\ref{ha1}) and (\ref{hp1}), the following relationships are
easily obtained
\begin{eqnarray}
\left({\partial\,{\cal H}_{\rm
A}(f_{_\ell})\over\partial\,f_{_\ell}}\right)^* ={1\over
T^*}\,\left[\mu^*-\left({\partial\,{\cal E}_{\rm A}\over\partial\,
f_{_\ell}}\right)^*\right] \ ,\label{h1}\\
\left({\partial\,{\cal
H}_p\left(I_{_{ij}}\right)\over\partial\,I_{_{ij}}}\right)^*
=-{\omega_{_{ij}}\over T^*}\left({\partial\,{\cal E}_{\rm
p}\over\partial\,I_{_{ij}}}\right)^* \ .\label{h2}
\end{eqnarray}
From Eq. (\ref{e}), by taking into account Eq.s
(\ref{h1})-(\ref{h2}) and particle conservation $\sum_\ell\int
f_{_\ell}\,d{\bfm v}=\sum_\ell\int f_{_\ell}^\ast\,d{\bfm v}$, one
easily realizes that
$E=E^*$ implies Eq. (\ref{h0}). $\bullet$\\

The main result with respect to stability can be summarized as
follows:\\

{\bf Proposition 2.} The functional $L$ is a strict Lyapounov
functional for the present dynamical system: the unique
equilibrium is asymptotically stable, and any initial state will
relax to it asymptotically in time.\\

{\it Proof}. First one easily proves that
\begin{eqnarray}
\frac{d\,L}{d\,t}={\cal D}\leq0 \ ,\label{dL}
\end{eqnarray}
with $d\,L/d\,t=0$ only at the unique equilibrium position. In
fact, from the definition (\ref{L}) it follows
\begin{equation}
\fl\hspace{20mm}\frac{d\,L}{d\,t}=\sum_\ell\int_{I\!\!R^3}\frac{\partial\,{\cal
H}_{\rm
A}(f_{_\ell})}{\partial\,f_{_\ell}}\,\frac{\partial\,f_{_\ell}}{\partial\,t}\,d{\bfm
v}+\sum_j\sum_{i<j}{1\over \omega_{_{ij}}
}\int_{S^2}\frac{\partial\,{\cal H}_{\rm
p}(I_{_{ij}})}{\partial\,I_{_{ij}}}\,\frac{\partial\,I_{_{ij}}}{\partial\,t}\,d{\bfm\Omega}
\ ,
\end{equation}
and taking into account Eq.s (\ref{ha1}) and (\ref{hp1}), from the
definition (\ref{dp}) follows Eq. (\ref{dL}).\\ On the other hand,
by taking into account Lemma 2, one has
\begin{eqnarray}
\nonumber\fl\hspace{5mm} L-L^*=H_{\rm A}-H_{\rm A}^\ast+H_{\rm
p}-H_{\rm
p}^\ast\\
\nonumber\fl\hspace{19mm}=\sum_\ell\int_{I\!\!R^3}{\cal H}_{\rm
A}(f_{_\ell})\,d{\bfm v} -\sum_\ell\int_{I\!\!R^3}{\cal H}_{\rm
A}(f_{_\ell}^\ast)\,d{\bfm v}\\
\nonumber\fl\hspace{24mm}+\sum_j\sum_{i<j}{1\over \omega_{_{ij}}
}\int_{S^2}{\cal H}_{\rm
p}(I_{_{ij}})\,d{\bfm\Omega}-\sum_j\sum_{i<j}{1\over
\omega_{_{ij}} }\int_{S^2}{\cal H}_{\rm
p}(I_{_{ij}}^\ast)\,d{\bfm\Omega}\\
\nonumber\fl\hspace{19mm}=\sum_\ell\int_{I\!\!R^3}{\cal H}_{\rm
A}(f_{_\ell})\,d{\bfm v} -\sum_\ell\int_{I\!\!R^3}{\cal H}_{\rm
A}(f_{_\ell}^\ast)\,d{\bfm
v}-\sum_\ell\int_{I\!\!R^3}\left({\partial\,{\cal H}_{\rm
A}(f_{_\ell})\over\partial\,f_{_\ell}}\right)^*
(f_{_\ell}-f_{_\ell}^*)\,d{\bfm v}\\
\nonumber\fl\hspace{24mm}+\sum_j\sum_{i<j}{1\over \omega_{_{ij}}
}\int_{S^2}{\cal H}_{\rm
p}(I_{_{ij}})\,d{\bfm\Omega}-\sum_j\sum_{i<j}{1\over
\omega_{_{ij}} }\int_{S^2}{\cal H}_{\rm
p}(I_{_{ij}}^\ast)\,d{\bfm\Omega}\\
\nonumber\fl\hspace{24mm}-\sum_j\sum_{i<j}{1\over \omega_{_{ij}}}
\int_{S^2}\left({\partial\,{\cal H}_{\rm
p}(I_{_{ij}})\over\partial\,I_{_{ij}}}
\right)^*(I_{_{ij}}-I_{_{ij}}^*)\,d{\bfm\Omega}\\
\fl\hspace{19mm}=\sum_\ell\int_{I\!\!R^3}\widehat{\cal H}_{\rm
A}(f_{_\ell})\,d{\bfm v}+\sum_j\sum_{i<j}{1\over \omega_{_{ij}}
}\int_{S^2}\widehat{\cal H}_{\rm p}(I_{_{ij}})\,d{\bfm\Omega} \ ,
\end{eqnarray}
where
\begin{eqnarray}
\widehat{\cal H}_{\rm A}(f_{_\ell})={\cal H}_{\rm
A}(f_{_\ell})-\left[{\cal H}_{\rm A}(f_{_\ell}^*)
+\left(\frac{\partial\,{\cal H}_{\rm A}(f_{_\ell})}
{\partial\,f_{_\ell}}\right)^*(f_{_\ell}-f_{_\ell}^*)\right] \ ,\label{hca}\\
\widehat{\cal H}_{\rm p}(I_{_{ij}})={\cal H}_{\rm
p}(I_{_{ij}})-\left[{\cal H}_{\rm p}(I_{_{ij}}^*)
+\left(\frac{\partial\,{\cal H}_{\rm p}(I_{_{ij}})}{\partial\,
I_{_{ij}}}\right)^*(I_{_{ij}}-I_{_{ij}}^*)\right] \ .\label{hcp}
\end{eqnarray}
Since both ${\cal H}_{\rm A}$ and ${\cal H}_{\rm p}$ are convex,
we have $\widehat{\cal H}_{\rm A}\ge0$ and $\widehat{\cal H}_{\rm
p}\ge0$, and then we can conclude that $L\ge L^\ast$ and attains
its minimum, $L=L^\ast$, only for the unique equilibrium position.
$\bullet$\\

In the next two Sections we inquire on equilibrium and stability
in two particular cases: photons in a background gas and atoms in
a background radiation.

%%%%%%%%%%%%%%%%%%%%%%%%%%%%%%%%%%%%%%%%%%%%%%%%%%%%%%%%%%%%%%%%%%%%%%%%%%%%%%%%%%%%%%
\sect{Photons in a background gas}

An usual assumption in radiation gasdynamics is that the
relaxation time due to atom-atom interactions is much quicker than
the one due to gas-radiation processes. Thus, in the kinetic
equation for photons we can fix the distribution of atoms as an
equilibrium function $f_{_\ell}^\ast$ at a certain temperature
$T$.

In order to study equilibrium and its stability, we introduce the
following functional
\begin{eqnarray}
{\cal C}_{\rm p}=\sum_j\sum_{i<j}{1\over
\omega_{_{ij}}}\int_{S^2}\ln\left[{\Phi_{_{ij}}\over
\Psi_{_{ij}}}\exp\left({\omega_{_{ij}} \over
T}\right)\right]\,\left({\partial\, I_{_{ij}}\over\partial\,t}
\right)^\diamond\,d{\bfm \Omega} \ ,\label{cp}
\end{eqnarray}
where $\diamond$ means that $f_{_\ell}$ has been substituted by
$f_{_\ell}^*$.\\After using the kinetic equation (\ref{kinetic})
Eq. (\ref{cp}) becomes
\begin{eqnarray}
{\cal C}_{\rm
p}=\sum_j\sum_{i<j}\int_{S^2}\ln\left[{\Phi_{_{ij}}\over
\Psi_{_{ij}}}\exp\left({\omega_{_{ij}} \over
T}\right)\right]\,\left[{\widetilde J}_{_{ij}}({\bfm
\Omega})\right]^\diamond\,d{\bfm \Omega} \ ,
\end{eqnarray}
and using Lemma 1 with all the $\gamma_{_\ell}({\bfm v})=0$ and
$\Gamma_{_{ij}}({\bfm\Omega})=\ln[(\Phi_{_{ij}}/\Psi_{_{ij}})\,\exp(\omega_{_{ij}}/T)]$
we obtain
\begin{eqnarray}
\nonumber\fl{\cal C}_{\rm
p}=\sum_j\sum_{i<j}\alpha_{_{ij}}\int_{S^2}\left[\varphi_{_j}^*({\bfm
v})\Psi_{_{ij}}({\bfm\Omega})\psi_{_i}^*({\bfm v})
-\varphi_{_i}^*({\bfm v})\Phi_{_{ij}}({\bfm\Omega})\psi_j^*({\bfm
v})\right]\ln\left[{\Phi_{_{ij}}\over
\Psi_{_{ij}}}\exp\left({\omega_{_{ij}} \over
T}\right)\right]d{\bfm v}d{\bfm \Omega} \ .\\
\end{eqnarray}
Finally, observing that
\begin{eqnarray}
\omega_{_{ij}}=\left({1\over2}\,m\,{\bfm
v}^2+E_{_j}-\mu\right)-\left({1\over2}\,m\,{\bfm
v}^2+E_{_i}-\mu\right) \ ,
\end{eqnarray}
and taking into account Eq. (\ref{a}), the quantity ${\cal C}_{\rm
p}$ becomes
\begin{eqnarray}
\nonumber {\cal C}_{\rm p}&=&\sum_j\sum_{i<j}\alpha_{_{ij}}
\int_{S^2}\left[\varphi_{_j}^*({\bfm
v})\,\Psi_{_{ij}}({\bfm\Omega})\,\psi_{_i}^*({\bfm v})
-\varphi_{_i}^*({\bfm
v})\,\Phi_{_{ij}}({\bfm\Omega})\,\psi_j^*({\bfm v})\right]\\
&\times&\ln{\varphi_{_i}^*({\bfm
v})\,\Phi_{_{ij}}({\bfm\Omega})\,\psi_{_j}^*({\bfm v})\over
\varphi_{_j}^*({\bfm
v})\,\Psi_{_{ij}}({\bfm\Omega})\,\psi_{_i}^*({\bfm
v})}\,d{\bfm\Omega} \ ,
\end{eqnarray}
which is manifestly not positive.\\ From usual arguments (see
proof of Proposition 1), the equilibrium condition $\partial\,
I_{_{ij}}/\partial\,t=0$ is equivalent to
\begin{eqnarray}
\varphi_{_j}^*({\bfm
v})\,\Psi_{_{ij}}({\bfm\Omega})\,\psi_{_i}^*({\bfm v})
=\varphi_{_i}^*({\bfm
v})\,\Phi_{_{ij}}({\bfm\Omega})\,\psi_{_j}^*({\bfm v}) \
.\label{56}
\end{eqnarray}
Again, after using Eq. (\ref{a}), from Eq. (\ref{56}) we find the
generalized Planck's law (\ref{b}) for photons.

In order to investigate the stability of such equilibrium, let's
introduce the following functional:
\begin{eqnarray}
L_{\rm p}=H_{\rm p}+{E_{\rm p}\over T} \ ,\label{Lp}
\end{eqnarray}
where $H_{\rm p}$ is defined in Eq. (\ref{hp}) and $E_{\rm p}$ is given in Eq. (\ref{ep}).\\

{\bf Proposition 3.} $L_{\rm p}$ is a Lyapounov functional for the
present problem.\\

{\it Proof.} First of all we find
\begin{eqnarray}
\frac{d\,L_{\rm p}}{d\,t}={\cal C}_{\rm p}\leq0 \ ,\label{c}
\end{eqnarray}
thus $L_{\rm p}$ is a decreasing function. In fact, from the
definition (\ref{Lp}) we have
\begin{eqnarray}
\fl\hspace{10mm}\frac{d\,L_{\rm p}}{d\,t}=\sum_j\sum_{i<j}{1\over
\omega_{_{ij}}}\int_{S^2} \,\left[\frac{\partial\,{\cal H}_{\rm
p}(I_{_{ij}})}{\partial\,I_{_{ij}}}+{\omega_{_{ij}}\over
T}\right]\,\left(\frac{\partial\,I_{_{ij}}}{\partial\,t}\right)^\diamond
\,d{\bfm\Omega} \ ,
\end{eqnarray}
and after using Eq. (\ref{hp1}) we obtain Eq. (\ref{c}).\\
Moreover, since
\begin{eqnarray}
\left({\partial\,{\cal H}_{\rm p}(I_{_{ij}})\over
\partial\,I_{_{ij}}}\right)^*
=-{\omega_{_{ij}}\over T} \ ,
\end{eqnarray}
as it follows from Eq.s (\ref{b}) and (\ref{hp1}), we can write
\begin{eqnarray}
\nonumber\fl\hspace{5mm}L_{\rm p}-L^*_{\rm p}=H_{\rm p}-H_{\rm
p}^\ast+{1\over
T}\left(E_{\rm p}-E_{\rm p}^\ast\right)\\
\nonumber\fl\hspace{5mm} =\sum_j\sum_{i<j}{1\over
\omega_{_{ij}}}\int_{S^2}{\cal H}_{\rm
p}(I_{_{ij}})\,d{\bfm\Omega}-\sum_j\sum_{i<j}{1\over
\omega_{_{ij}}}\int_{S^2}{\cal H}_{\rm
p}(I_{_{ij}}^\ast)\,d{\bfm\Omega}+{1\over
T}\sum_j\sum_{i<j}\int_{S^2}\left(I_{_{ij}}-I_{_{ij}}^\ast\right)\,d{\bfm\Omega}\\
\nonumber\fl\hspace{5mm} =\sum_j\sum_{i<j}{1\over
\omega_{_{ij}}}\int_{S^2}\left[{\cal H}_{\rm p}(I_{_{ij}})-{\cal
H}_{\rm p}(I_{_{ij}}^\ast)-\left(\frac{\partial\,{\cal H}_{\rm
p}(I_{_{ij}})}{\partial\,I_{_{ij}}}\right)^\ast\left(I_{_{ij}}-
I_{_{ij}}^\ast\right)\right]\,d{\bfm\Omega}\\
\fl\hspace{5mm} =\sum_j\sum_{i<j}{1\over \omega_{_{ij}}}\int_{S^2}
\widehat{\cal H}_{\rm p}(I_{_{ij}}) \,d{\bfm\Omega} \ ,
\end{eqnarray}
where ${\widehat{\cal H}}_{\rm p}$ is given in Eq. (\ref{hcp}).
Because ${\cal H}_{\rm p}$ is convex we conclude that $L_{\rm
p}\geq L_{\rm p}^\ast$, that is $L_{\rm p}$ attains a minimum at
equilibrium. $\bullet$\\

Taking into account of the definition of $L_{\rm p}$ it is easy to
realize that Eq. (\ref{c}) is equivalent to Clausius inequality
\begin{eqnarray}
\frac{d\,S_{\rm p}}{d\,t}\geq {1\over T}\,\frac{d\,E_{\rm
p}}{d\,t} \ ,
\end{eqnarray}
where $S_{\rm p}=-H_{\rm p}$. From $L_{\rm p}-L_{\rm
p}^\ast\geq0$ follows that the quantity ${\cal S}_{\rm p}=S_{\rm
p}-E_{\rm p}/T$ is always increasing and attains a maximum at
equilibrium.
%%%%%%%%%%%%%%%%%%%%%%%%%%%%%%%%%%%%%%%%%%%%%%%%%%%%%%%%%%%%%%%%%%%%%%%%%%%%%%%%%%%%%%
\sect{Atoms in a background radiation}

Suppose now that temperature is high enough so that we can
consider photons as an equilibrium  background. This means that,
in the kinetic equations for atoms, we can fix the distribution of
photons as an equilibrium function $I_{_{ij}}^\ast$ at a certain
temperature $\cal T$.

In order to study equilibrium and its stability, we introduce the
following functional
\begin{eqnarray}
{\cal C}_{\rm
A}=\sum_\ell\int_{I\!\!R^3}\ln\left\{{\varphi_{_\ell}\over\psi_{_\ell}}
\,\exp\left[{1\over T}\left({1\over2}\,m\,{\bfm v}^2+E_{_\ell}-\mu
\right)\right]\right\}\,\left({\partial\,f_{_\ell}\over\partial\,
t}\right)^\sharp\,d{\bfm v} \ ,\label{cca}
\end{eqnarray}
where $\sharp$ means that $I_{_{ij}}$ has been substituted by
$I_{_{ij}}^\ast$. After using the kinetic equation
(\ref{boltzmann}), Eq. (\ref{cca}) becomes
\begin{eqnarray}
\fl\hspace{10mm}{\cal C}_{\rm
A}=\sum_\ell\int_{I\!\!R^3}\ln\left\{{\varphi_{_\ell}\over\psi_{_\ell}}
\,\exp\left[{1\over T}\left({1\over2}\,m\,{\bfm v}^2+E_{_\ell}-\mu
\right)\right]\right\}\,\left[J_{_\ell}({\bfm
v})+\widetilde{J}_{_\ell}({\bfm v})\right]^\sharp\,d{\bfm v} \
.\label{ca3}
\end{eqnarray}
Using Lemma 1 with $\gamma_{_\ell}({\bfm
v})=\ln\{(\varphi_{_\ell}/\psi_{_\ell})\,\exp\left[\left(m\,{\bfm
v}^2/2+E_{_\ell}-\mu \right)/T\right]\}$ and all
$\Gamma_{ij}({\bfm\Omega})=0$, Eq. (\ref{ca3}) becomes
\begin{eqnarray}
\nonumber \fl\hspace{10mm}{\cal C}_{\rm A}
=\sum_\ell\sum_j\sum_{i\leq j} \int_{I\!\!R^3\times I\!\!R^3\times
S^2} g\,\sigma_{_{\ell j}}^{^{\ell i
}}(g,\,\zeta)\\
\nonumber \fl\hspace{16mm}\times[\varphi_{_\ell}({\bfm
v}_{_{ij}}^+)\,\varphi_{_i}({\bfm
w}_{_{ij}}^+)\,\psi_{_\ell}({\bfm v})\,\psi_{_j}({\bfm
w})-\varphi_{_\ell}({\bfm v})\,\varphi_{_j}({\bfm
w})\psi_{_\ell}({\bfm v}_{_{ij}}^+)\,\psi_{_i}({\bfm
w}_{_{ij}}^+)]\\
\nonumber \fl\hspace{16mm}\times\ln\left\{{\varphi_{_\ell}({\bfm
v})\,\varphi_{_j}({\bfm w})\,\psi_{_\ell}({\bfm
v}_{_{ij}}^+)\,\psi_{_i}({\bfm w}_{_{ij}}^+)\over
\varphi_{_\ell}({\bfm v}_{_{ij}}^+) \,\varphi_{_i}({\bfm
w}_{_{ij}}^+)\,\psi_{_\ell}({\bfm v})\,\psi_{_j}({\bfm
w})}\right.\\
\nonumber
\fl\hspace{16mm}\left.\times\exp\left[{E_{_j}-E_{_i}\over
T}-{m\over2\,T}\left(({\bfm v}_{_{ij}}^+)^2+({\bfm
w}_{_{ij}}^+)^2-{\bfm v}^2-{\bfm w}^2\right)\right]\right\}\,d{\bfm v}\,d{\bfm w}\,d{\bfm\Omega}^\prime\\
\nonumber\fl\hspace{16mm}+\sum_j\sum_{i<j}\alpha_{_{ij}}\int_{I\!\!R^3\times
S^2} [\varphi_{_j}({\bfm
v})\,\Psi_{_{ij}}^*({\bfm\Omega})\,\psi_{_i}({\bfm
v})-\varphi_{_i}({\bfm
v})\,\Phi_{_{ij}}^*({\bfm\Omega})\,\psi_{_j}({\bfm
v})]\\
\fl\hspace{16mm}\times\ln\left[{\varphi_{_i}({\bfm
v})\,\psi_{_j}({\bfm v})\over \varphi_{_j}({\bfm
v})\,\psi_{_i}({\bfm v})}\,\exp\left(-{E_{_j}-E_{_i}\over
T}\right)\right]\,d{\bfm v}\,d{\bfm\Omega} \ ,\label{ca2}
\end{eqnarray}
and taking into account the energy conservation and Eq. (\ref{b})
\begin{eqnarray}
\omega_{_{ij}}=E_{_j}-E_{_i}= {1\over2}\,m\left[({\bfm
v}_{_{ij}}^+)^2+({\bfm w}_{_{ij}}^+)^2-{\bfm v}^2-{\bfm
w}^2\right] \ ,
\end{eqnarray}
we obtain
\begin{eqnarray}
\nonumber \fl\hspace{10mm}{\cal C}_{\rm A}
=\sum_\ell\sum_j\sum_{i\leq j} \int_{I\!\!R^3\times I\!\!R^3\times
S^2} g\,\sigma_{_{\ell j}}^{^{\ell i
}}(g,\,\zeta)\\
\nonumber \fl\hspace{16mm}\times[\varphi_{_\ell}({\bfm
v}_{_{ij}}^+)\,\varphi_{_i}({\bfm
w}_{_{ij}}^+)\,\psi_{_\ell}({\bfm v})\,\psi_{_j}({\bfm
w})-\varphi_{_\ell}({\bfm v})\,\varphi_{_j}({\bfm
w})\psi_{_\ell}({\bfm v}_{_{ij}}^+)\,\psi_{_i}({\bfm
w}_{_{ij}}^+)]\\
\nonumber \fl\hspace{16mm}\times\ln{\varphi_{_\ell}({\bfm
v})\,\varphi_{_j}({\bfm w})\,\psi_{_\ell}({\bfm
v}_{_{ij}}^+)\,\psi_{_i}({\bfm w}_{_{ij}}^+)\over
\varphi_{_\ell}({\bfm v}_{_{ij}}^+) \,\varphi_{_i}({\bfm
w}_{_{ij}}^+)\,\psi_{_\ell}({\bfm v})\,\psi_{_j}({\bfm
w})}\,d{\bfm v}\,d{\bfm w}\,d{\bfm\Omega}^\prime\\
\nonumber\fl\hspace{16mm}+\sum_j\sum_{i<j}\alpha_{_{ij}}\int_{I\!\!R^3\times
S^2} [\varphi_{_j}({\bfm
v})\,\Psi_{_{ij}}^*({\bfm\Omega})\,\psi_{_i}({\bfm
v})-\varphi_{_i}({\bfm
v})\,\Phi_{_{ij}}^*({\bfm\Omega})\,\psi_{_j}({\bfm
v})]\\
\fl\hspace{16mm}\times\ln{\varphi_{_i}({\bfm
v})\,\Phi_{_{ij}}^*({\bfm\Omega})\,\psi_{_j}({\bfm v})\over
\varphi_{_j}({\bfm
v})\,\Psi_{_{ij}}^*({\bfm\Omega})\,\psi_{_i}({\bfm v})}\,d{\bfm
v}\,d{\bfm\Omega} \ ,\label{ca1}
\end{eqnarray}
which is manifestly non positive.\\
Following the same steps given in Proposition 1, we obtain that
the equilibrium condition $\partial f_{_\ell}/\partial t=0$ is
equivalent to
\begin{eqnarray}
\varphi_{_\ell}({\bfm v}_{_{ij}}^+)\,\varphi_{_i}({\bfm
w}_{_{ij}}^+)\,\psi_{_\ell}({\bfm v})\,\psi_{_j}({\bfm
w})=\varphi_{_\ell}({\bfm v})\,\varphi_{_j}({\bfm
w})\psi_{_\ell}({\bfm v}_{_{ij}}^+)\,\psi_{_i}({\bfm w}_{_{ij}}^+)
 \ ,\label{1}\\
\hspace{15mm}\varphi_{_j}({\bfm
v})\,\Psi_{_{ij}}^*({\bfm\Omega})\,\psi_{_i}({\bfm
v})=\varphi_{_i}({\bfm
v})\,\Phi_{_{ij}}^*({\bfm\Omega})\,\psi_{_j}({\bfm v}) \
.\label{2}
\end{eqnarray}
The first Eq. (\ref{1}) gives
\begin{eqnarray}
\ln{\varphi_{_\ell}\over\psi_{_\ell}}=-{1\over
T}\,\left(\frac{1}{2}\,m\,{\bfm v}^2+E_{_\ell}-\mu\right) \
,\label{ppp}
\end{eqnarray}
while the second Eq. (\ref{2}), after using Eq. (\ref{ppp}) gives
$T={\cal T}$.

 In order to study the stability of this equilibrium
we introduce the following functional
\begin{eqnarray}
L_{\rm A}=H_{\rm A}+{E_{\rm A}\over T} \ ,\label{La}
\end{eqnarray}
with $H_{\rm A}$ and $E_{\rm A}$ given in Eq.s (\ref{ha}) and
(\ref{ea}), respectively.\\

{\bf Proposition 4.} $L_{\rm A}$ is a Lyapounov functional for the
present problem.\\

{\it Proof.} First we find
\begin{eqnarray}
\frac{d\,L_{\rm A}}{d\,t}={\cal C}_{\rm A}\leq0 \ ,\label{ca}
\end{eqnarray}
so that $L_{\rm A}$ is a decreasing function. In fact, from the
definition (\ref{La}) we obtain
\begin{eqnarray}
\fl\hspace{10mm}\frac{d\,L_{\rm
A}}{d\,t}=\sum_\ell\int_{I\!\!R^3}\left[\frac{\partial\,{\cal
H}_{\rm A}(f_{_\ell})}{\partial\,f_{_\ell}}+{1\over
T}\left({1\over2}\,m\,{\bfm
v}^2+E_{_\ell}\right)\right]\,\left(\frac{\partial\,f_{_\ell}}{\partial\,t}\right)^\sharp
\,d{\bfm v} \ .
\end{eqnarray}
After using Eq. (\ref{ha1}) and taking into account for particle
conservation we obtain the relation
(\ref{ca}).\\
Moreover, since
\begin{eqnarray}
\left({\partial\,{\cal H}_{\rm A}(f_{_\ell})\over\partial\,
f_{_\ell}}\right)^*=-\frac{1}{T}\,\left({1\over2}\,m\,{\bfm
v}^2+E_{_\ell}-\mu\right) \ ,
\end{eqnarray}
as it follows from Eq.s (\ref{a}) and (\ref{ha1}), by accounting
for particle and total energy conservation we can write
\begin{eqnarray}
\nonumber\fl L_{\rm A}-L^*_{\rm A}=H_{\rm A}-H_{\rm
A}^\ast+{1\over
T}\left(E_{\rm A}-E_{\rm A}^\ast\right)\\
\nonumber\fl=\sum_\ell\int_{I\!\!R^3}{\cal H}_{\rm
A}(f_{_\ell})\,d{\bfm v}-\sum_\ell\int_{I\!\!R^3}{\cal H}_{\rm
A}(f_{_\ell}^\ast)\,d{\bfm v}+{1\over
T}\sum_\ell\int_{I\!\!R^3}\left({1\over2}\,m\,{\bfm
v}^2+E_{_\ell}-\mu\right)\,
\left(f_{_\ell}-f_{_\ell}^\ast\right)\,d{\bfm v}\\
\nonumber\fl=\sum_\ell\int_{I\!\!R^3}\left[{\cal H}_{\rm
A}(f_{_\ell})-{\cal H}_{\rm
A}(f_{_\ell}^\ast)-\left(\frac{\partial\,{\cal
H}_{\rm A}(f_{_\ell})}{\partial\,f_{_\ell}}\right)^\ast\left(f_{_\ell}-f_{_\ell}^\ast\right)\right]\,d{\bfm v}\\
\fl=\sum_\ell\int_{I\!\!R^3} \widehat{\cal H}_{\rm A}(f_{_\ell})
\,d{\bfm v} \ ,
\end{eqnarray}
with ${\widehat{\cal H}}_{\rm A}$ given in Eq. (\ref{hca}). Due to
the convexity of ${\cal H}_{\rm A}$ we can conclude that $L_{\rm
A}\geq L_{\rm A}^\ast$ and
$L_{\rm A}$ attains a minimum at equilibrium.  $\bullet$\\

Finally we can verify that Eq. (\ref{ca}) is equivalent to
Clausius inequality
\begin{eqnarray}
\frac{d\,S_{\rm A}}{d\,t}\geq{1\over T}\frac{d\,E_{\rm A}}{d\,t} \
,
\end{eqnarray}
where $S_{\rm A}=-H_{\rm A}$. From $L_{\rm A}-L_{\rm
A}^\ast\geq0$ follows that the quantity ${\cal S}_{\rm A}=S_{\rm
A}-E_{\rm A}/T$ is always increasing and attains a maximum at
equilibrium.

%%%%%%%%%%%%%%%%%%%%%%%%%%%%%%%%%%%%%%%%%%%%%%%%%%%%%%%%%%%%%%%%%%%%%%%%%%%%%%%%%%%%%%
\sect{Numerical Simulations}

For the purpose of numerical calculations, we consider a
homogeneous and  isotropic case and keep the generalization for
photons only. Atoms are treated as classical particles. This means
that substitution (\ref{clas}) is performed, while (\ref{clas1})
is not applied. A couple of different cases are considered for the
functions $\Phi$ and $\Psi$.\\ By integrating Eq.
(\ref{boltzmann}) over $d{\bfm v}$ we obtain
\begin{eqnarray}
{d\,N_{_\ell}\over d\,t}=\sum_{i>\ell}(S_{_{\ell
i}}+4\,\pi\,B_{_{\ell
i}})-\sum_{i<\ell}(S_{_{i\ell}}+4\,\pi\,B_{_{i\ell}}) \ ,\label{n}
\end{eqnarray}
where
\begin{eqnarray}
N_{_\ell}=\int_{I\!\!R^3} f_{_\ell}({\bfm v})\,d{\bfm v} \ ,
\end{eqnarray}
is the number density of atoms at level $\ell$ and
\begin{eqnarray}
\fl\hspace{10mm}S_{_{ij}}=\sum_\ell\int_{I\!\!R^3\times
I\!\!R^3\times S^2} g\,\sigma_{_{\ell j}}^{^{\ell
i}}(g,\,\zeta)[f_{_\ell}({\bfm v})\,f_{_j}({\bfm
w})-f_{_\ell}({\bfm
v}^+_{_{ij}})\,f_{_i}({\bfm w}^+_{_{ij}})]\,d{\bfm v}\,d{\bfm w}\,d{\bfm\Omega}^\prime \ ,\\
\fl\hspace{10mm}B_{_{ij}}=\alpha_{_{ij}}[N_{_j}\,\Psi(I_{_{ij}})-
N_{_i}\,\Phi(I_{_{ij}})] \ ,
\end{eqnarray}
are the source terms due to the elastic/inelastic and
emission/absorption interactions, respectively.\\ By multiplying Eq.
(\ref{boltzmann}) by $m\,{\bfm v}^2/2$, by summing over $\ell$
and by integrating over $d{\bfm v}$ we obtain
\begin{eqnarray}
{3\over 2}\,N\,{d\,T\over
d\,t}=\sum_j\sum_{i<j}\omega_{_{ij}}\,S_{_{ij}} \ ,\label{t}
\end{eqnarray}
with the total number density of the mixture $N=\sum_\ell
N_{_\ell}$ a constant, and temperature given by
\begin{eqnarray}
T={1\over 3\,N}\,{\rm Tr}\,I\!\!P \ ,
\end{eqnarray}
where
\begin{eqnarray}
I\!\!P=m\,\sum_\ell\int_{I\!\!R^3}\left({\bfm v}\otimes{\bfm
v}\right)\,f_{_\ell}({\bfm v})\,d{\bfm v} \ ,
\end{eqnarray}
is the stress tensor.\\
Differently, by integrating Eq. (\ref{kinetic}) over
$d{\bfm\Omega}$ follows
\begin{eqnarray}
{d\,I_{_{ij}}\over d\,t}=\omega_{_{ij}}\,B_{_{ij}} \ .\label{i}
\end{eqnarray}
Suppose now that elastic atom-atom interactions prevail over the
inelastic and gas-radiation ones (consistently with the usual
local thermodynamic equilibrium approximation). Thus, according to
the zero-order Chapmann-Enskog approximation we can calculate
$S_{_{ij}}$ by means of the Maxwellians
\begin{eqnarray}
f_{_\ell}({\bfm v})=N_{_\ell}\left({m\over 2\,\pi\,
T}\right)^{3/2}\,\exp\left(-{m\,{\bfm v}^2\over 2\,T}\right) \ .
\end{eqnarray}
It is found
\begin{eqnarray}
S_{_{ij}}=G_{_{ij}}\left[N_{_j}-N_{_i}\,\exp(-\frac{\omega_{_{ij}}}{
T})\right] \ ,
\end{eqnarray}
with
\begin{eqnarray}
\fl\hspace{10mm}G_{_{ij}}=\sum_\ell N_{_\ell}\,\gamma_{_{\ell j}}^{^{\ell i}}(T) \ ,\\
\fl\hspace{10mm}\gamma_{_{\ell j}}^{^{\ell i}}(T)=\left({m\over
2\,\pi\,T}\right)^3 \int_{I\!\!R^3\times I\!\!R^3\times S^2}
g\,\sigma_{_{\ell j}}^{^{\ell i}}(g,\,\chi)\,\exp\left[ -{m\over
2\,T}\,({\bfm v}^2+{\bfm w}^2)\right]\,d{\bfm v}\,d{\bfm
w}\,d{\bfm \Omega}^\prime \ .
\end{eqnarray}
Remark that the equilibrium condition is equivalent to
$S_{_{ij}}=B_{_{ij}}=0$, from which follows the relation
\begin{eqnarray}
N_{_i}\,\exp\left({E_{_i}\over
T}\right)=N_{_j}\,\exp\left({E_{_j}\over T}\right) \ .
\end{eqnarray}
Taking into account the total number conservation, we obtain the
atoms distribution function
\begin{eqnarray}
N_{_\ell}=\frac{N}{Z}\,\exp\left(-\frac{E_{_\ell}}{T}\right) \
,\label{nl}
\end{eqnarray}
with
\begin{eqnarray}
Z=\sum_\ell\exp\left(-\frac{E_{_\ell}}{T}\right) \ .\label{nt}
\end{eqnarray}
We observe that Eq.s (\ref{nl}) and (\ref{nt}) assume the standard
expression for the equilibrium distributions of the atomic levels.
Notwithstanding, because the generalized statistics is kept for
the photons, the equilibrium temperature is different respect to
the case in which also photons obey to the standard Bose-Einstein
statistics. Consequently, the values of the population levels
$N_{_\ell}$ are numerically different.

Eq.s (\ref{n}), (\ref{t}) and (\ref{i}) constitute a close system
for the unknown quantities $N_{_\ell}$, $T$ and $I_{_{ij}}$. They
can be integrated by means of numerical calculation, with suitable initial conditions.\\

\begin{figure}[h]
\begin{center}
\includegraphics[width=1\textwidth]{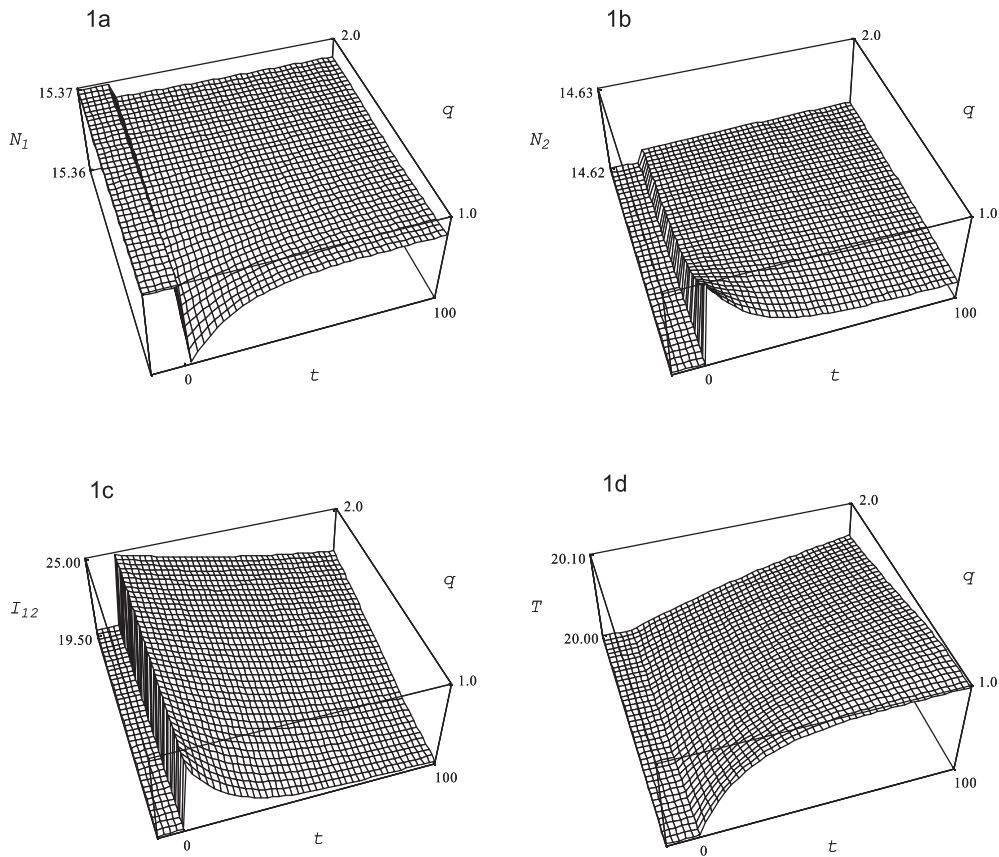}
\caption{Plot (in arbitrary units) of time evolution of $N_{_1}$
(1a), $N_{_2}$ (1b), $I_{_{12}}$ (1c) and $T$ (1d) v.s. $q$, after
injection, at $t=0$, of a flash of photons obeying to the $q$-Bose
statistics.\label{fig1} }
\end{center}
\vspace{10mm}
\end{figure}
\begin{figure}[h]
\begin{center}
\includegraphics[width=1\textwidth]{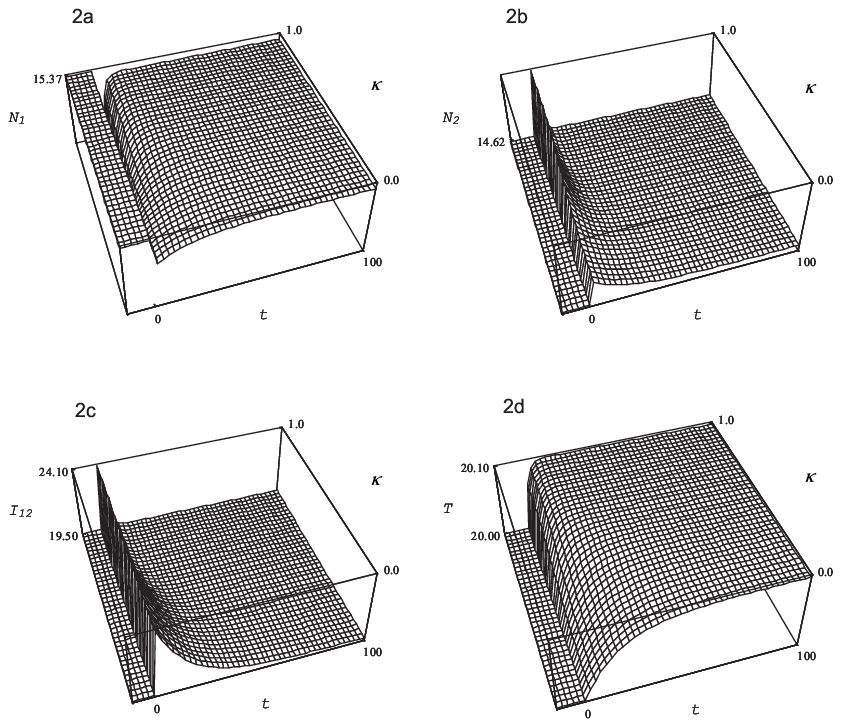}
\caption{Plot (in arbitrary units) of time evolution of $N_{_1}$
(2a), $N_{_2}$ (2b), $I_{_{12}}$ (2c) and $T$ (2d) v.s. $\kappa$,
after injection, at $t=0$, of a flash of photons obeying to the
$\kappa$-Bose statistics. \label{fig2} }
\end{center}
\vspace{10mm}
\end{figure}

In the following, we consider a system of $N$ atoms with $n=2$
levels and monochromatic photons with frequency $\omega_{_{12}}$.
The system is initially in the equilibrium configuration at
temperature $T^\ast$. From condition $B_{_{12}}=0$ follows the
distribution $I_{_{12}}^\ast$.

Two different simulation are take into account. In the first
simulation, a flash of photons is injected at the time $t=0$, from
an isotropic external source at temperature $T^\prime$. The
relaxation of the system to the new equilibrium configuration is
depicted v.s. time in the following figures 1 and 2, concerning
the results of two different generalizations adopted for the
photons and described in the Appendix B. They take into account
for an asymptotic inverse power law decay of the distribution
function $I_{_{12}}$
with respect to $\omega_{ij}$.\\
First we adopt the model proposed by
B\"uy\"ukkili\c c et al. \cite{5,6}. The time evolution of the
physical meaningful quantities $N_{_1}$, $N_{_2}$, $I_{_{12}}$ and
$T$ v.s. the deformed parameter $q$ is depicted in figures (1a), (1b), (1c) and (1d),
respectively.\\
Then, we adopt the model proposed by Kaniadakis \cite{11}. The
time evolution of the physical meaningful quantities $N_{_1}$,
$N_{_2}$, $I_{_{12}}$ and $T$ v.s. the deformed parameter $\kappa$
is depicted in figure (2a), (2b), (2c) and (2d), respectively.

In the second simulation, a constant and isotropic external photon
source is inserted at the time $t=0$. Again, both the
$q$-deformation and the $\kappa$-deformation are considered. The
time evolution of the same physical quantities $N_{_1}$, $N_{_2}$,
$I_{_{12}}$ and $T$ v.s. the deformed parameter $q$ [figures (3a),
(3b), (3c) and (3d)] and v.s. the deformed parameter $\kappa$
[figures (4a), (4b), (4c) and (4d)] are plotted.
\begin{figure}[h]
\begin{center}
\includegraphics[width=1\textwidth]{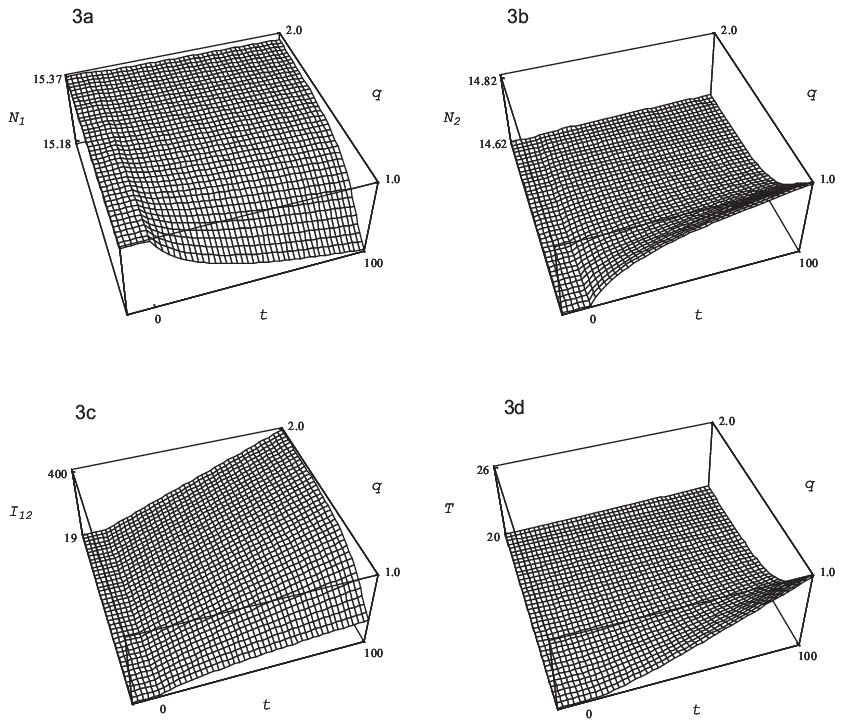}
\caption{Plot (in arbitrary units) of time
evolution of $N_{_1}$ (3a), $N_{_2}$ (3b), $I_{_{12}}$ (3c) and
$T$ (3d) v.s. $q$, after the introduction, at $t=0$, of a constant
photons source obeying to the $q$-Bose statistics. \label{fig3} }
\end{center}
\vspace{10mm}
\end{figure}
\begin{figure}[h]
\begin{center}
\includegraphics[width=1\textwidth]{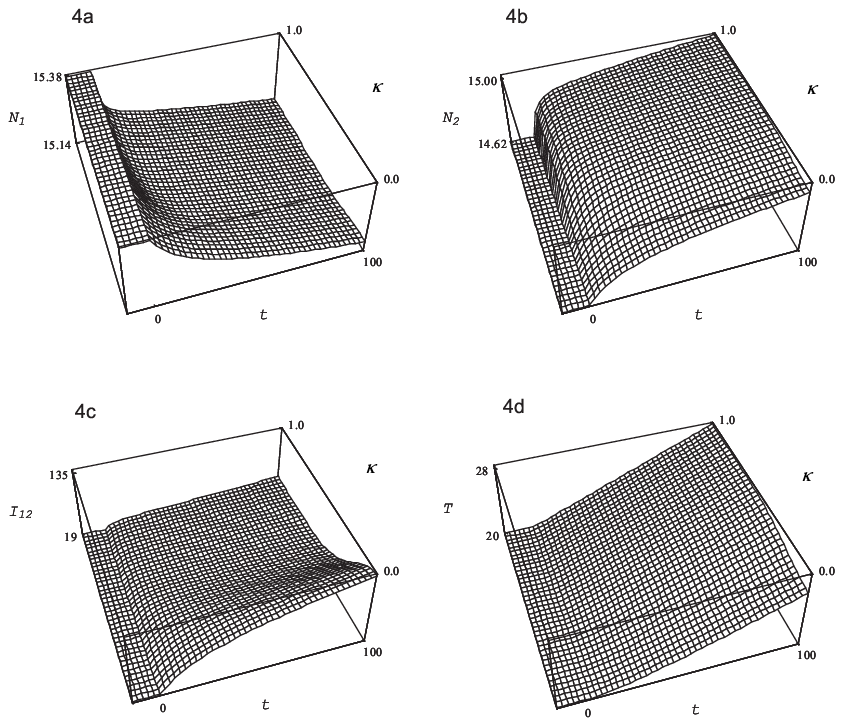}
\caption{Plot (in arbitrary units) of time
evolution of $N_{_1}$ (4a), $N_{_2}$ (4b), $I_{_{12}}$ (4c) and
$T$ (4d) v.s. $\kappa$, after the introduction, at $t=0$, of a
constant photons source obeying to the $\kappa$-Bose statistics.
\label{fig4} }
\end{center}
\vspace{10mm}
\end{figure}
 A comparison
between the $q$-deformed case and the $\kappa$-deformed case leads
to the following considerations. With the respect to the classical
case, the more the system is $\kappa$-deformed, the steeper the
evolution is. On the contrary,
the more the system is $q$-deformed the softer the evolution is.\\
We remark that this opposite behavior between the $q$-deformed
case and the $\kappa$-deformed case is a consequence of the choice
$q>1$. Such a choice arises from the observation that, in the
present problem, for values of $q<1$ a cut-off in the energy
spectrum is required.
(see Eq. (\ref{expq}) in the Appendix B).\\
Moreover, from the second simulation we observe that, as
well-known for the non-deformed case, in presence of a constant
pumping of photons, the two number densities reach asymptotically,
as $t\to\infty$, to the same value. Such a
feature is preserved also in the deformed cases but with a different time-scale.\\

%%%%%%%%%%%%%%%%%%%%%%%%%%%%%%%%%%%%%%%%%%%%%%%%%%%%%%%%%%%%%%%%%%%%%%%%%%%%%%%%%%%%%%
\sect{Conclusions} In order to improve the knowledge of the
generalized statistical mechanics and generalized kinetic theory,
we have investigated, in the present paper, a system of atoms and
photons obeying to a very general statistics. The kinetic
equations has been obtained, in the extended Boltzmann picture,
through the introduction of characteristic departure and arrival
functions for atoms and photons. In the space homogeneous case we
have studied the equilibrium configuration. Such equilibrium is
given by a modified Planck's law for the radiation and
by a generalized distribution function for the atoms.\\
By means of Lyapounov's theory we have studied on the stability of
this unique equilibrium configuration which maximize the entropic
functional of the system.

In the second part of this paper we have considered a homogeneous
and isotropic case keeping the generalization for photons only.
Atoms are treated as classical particles. Under suitable
assumptions on the relaxation times of the various interaction
processes, according to the zero-order Chapmann-Enskog
approximation, we have obtained a close system of macroscopic
equations for the unknown dynamical quantities: the number density
of atoms $N_{_\ell}$ at energy $E_{_\ell}$, the temperature of the
system $T$ and the
intensity $I_{_{ij}}$ at frequency $\omega_{_{12}}$.\\
We have shown the results of some numerical simulations for a
system with $n=2$ energy levels. In the first simulation the
system, initially at equilibrium at the temperature $T^\ast$, is
perturbed by a flash of photons injected by a external source at a
different temperature $T^\prime$. The relaxation of the system to
the new equilibrium state for the physically relevant quantities
$N_{_1}$, $N_{_2}$, $T$ and $I_{_{12}}$ has been plotted. In the
second simulation the system, initially at equilibrium at the
temperature $T^\ast$, is disturbed by a constant source of
photons. The time evolution of the same physically
relevant quantities has been plotted.\\

%%%%%%%%%%%%%%%%%%%%%%%%%%%%%%%%%%%%%%%%%%%%%%%%%%%%%%%%%%%%%%%%%%%%%%%%%%%%
\appendix
\sect{} For sake of completeness, we report in this Appendix on
the classical kinetic equations for atoms and photons, as known in
literature.\\
The distribution function of atoms $A_\ell$ obeys to the following
system of Boltzmann equations
\begin{eqnarray}
\frac{\partial\,f_{_\ell}}{\partial\,t}+{\bfm
v}\cdot{\bfm\nabla}f_{_\ell}=J_{_\ell}({\bfm v})+{\widetilde
J}_{_\ell}({\bfm v}) \ ,\label{boltzmanna}
\end{eqnarray}
where in the right hand side of Eq. (\ref{boltzmanna})
$J_{_\ell}({\bfm v})$ and ${\widetilde J}_{_\ell}({\bfm v})$
describe the contribution due to the atom-atom interaction and the
contribution due to the atom-photon
interaction, respectively.\\
In detail, for elastic/inelastic interactions between atoms we
have
\begin{eqnarray}
\fl\hspace{8mm}J_{_\ell}({\bfm v})=\sum_j\sum_{i\leq
j}J^{^{(1)}}_{_{ij\ell}}({\bfm v})+\sum_j\sum_{i\leq
j}J^{^{(2)}}_{_{ij\ell}}({\bfm v})+\sum_k\sum_{j\geq
\ell}J^{^{(3)}}_{_{jk\ell}}({\bfm v})+\sum_k\sum_{i\leq
\ell}J^{^{(4)}}_{_{ik\ell}}({\bfm v}) \ ,\label{atomsa}
\end{eqnarray}
with
\begin{eqnarray}
\fl\hspace{10mm}J^{^{(1)}}_{_{ij\ell}}({\bfm
v})=\int_{I\!\!R^3\times S^2} g\,\sigma_{_{\ell j}}^{^{\ell
i}}(g,\,\zeta)\,\left[f_{_\ell}({\bfm v}^+_{_{ij}})\,f_{_i}({\bfm
w}^+_{_{ij}}) -f_{_\ell}({\bfm v})\,f_{_j}({\bfm
w})\right]\,d{\bfm w}\,d{\bfm\Omega}^\prime \ ,\label{j1a}
\end{eqnarray}
\begin{eqnarray}
\fl\hspace{10mm}J^{^{(2)}}_{_{ij\ell}}({\bfm
v})=\int_{I\!\!R^3\times S^2} g\,\sigma_{_{\ell i}}^{^{\ell
j}}(g,\,\zeta)\left[f_{_\ell}({\bfm v}^-_{_{ij}})\,f_{_j}({\bfm
w}^-_{_{ij}})-f_{_\ell}({\bfm v})\,f_{_i}({\bfm w})
\right]\,d{\bfm w}\,d{\bfm\Omega}^\prime \ ,\label{j2a}
\end{eqnarray}
\begin{eqnarray}
\fl\hspace{10mm}J^{^{(3)}}_{_{jk\ell}}({\bfm
v})=\int_{I\!\!R^3\times S^2} g\,\sigma_{_{k\ell
}}^{^{kj}}(g,\,\zeta)\left[f_{_j}({\bfm v}^-_{_{\ell
j}})\,f_{_k}({\bfm w}^-_{_{\ell j}}) -f_{_\ell}({\bfm
v})\,f_{_k}({\bfm w})\right]\,d{\bfm w}\,d{\bfm\Omega}^\prime \
,\label{j3a}
\end{eqnarray}
\begin{eqnarray}
\fl\hspace{10mm}J^{^{(4)}}_{_{ik\ell}}({\bfm
v})=\int_{I\!\!R^3\times S^2}
g\,\sigma_{_{k\ell}}^{^{ki}}(g,\,\zeta)\left[f_{_i}({\bfm
v}^+_{_{i\ell}})\,f_{_k}({\bfm w}^+_{_{i\ell}})-f_{_\ell}({\bfm
v})\,f_{_k}({\bfm w})\right]\,d{\bfm w}\,d{\bfm\Omega}^\prime \
,\label{j4a}
\end{eqnarray}
where $\sigma_{_{\ell i}}^{^{\ell j}}$ and $\sigma_{_{\ell
j}}^{^{\ell i}}$ are the cross sections for forward and backward
reactions, describing elastic and inelastic interactions.
\\ In Eq.s
(\ref{j1a})-(\ref{j4a}) we have
\begin{eqnarray} {\bfm
v}^\pm_{_{ij}}={1\over2}\,({\bfm v}+{\bfm
w}+g_{_{ij}}^\pm\,{\bfm\Omega}^\prime) \ ,\hspace{10mm}{\bfm
w}^\pm_{_{ij}}={1\over2}\,({\bfm v}+{\bfm
w}-g_{_{ij}}^\pm\,{\bfm\Omega}^\prime) \ ,
\end{eqnarray}
\begin{eqnarray}
g_{_{ij}}^\pm=\sqrt{g^2\pm{4\over m}\,(E_{_j}-E_{_i})} \
,\hspace{8mm}g=|{\bfm v}-\bfm{w}| \ ,
\end{eqnarray}
\begin{eqnarray}
{\bfm\Omega}=\frac{1}{g}\,({\bfm v}-{\bfm w}) \
,\hspace{27mm}\cos\zeta={\bfm\Omega}\cdot{\bfm\Omega}^\prime \ ,
\end{eqnarray}
and the two-dimensional unit sphere $S^2$ is the domain of
integration for the unit vector ${\bfm\Omega}^\prime$. The four
contributions to $J_{_\ell}({\bfm v})$ correspond to the cases in
which $A_{_\ell}$ plays the role, in the reaction scheme
(\ref{re}), of $A_{_k}$ in the r.h.s., $A_{_k}$ in the l.h.s.,
$A_{_i}$ and $A_{_j}$, respectively.\\
For gas-radiation interactions, one can write
\begin{eqnarray}
{\widetilde J}_{_\ell}({\bfm v})=\sum_{i>\ell}\int_{S^2}{\widehat
J}_{_{\ell i}}({\bfm
v},\,{\bfm\Omega})\,d{\bfm\Omega}-\sum_{i<\ell}\int_{S^2}{\widehat
J}_{_{i\ell}}({\bfm v},\,{\bfm\Omega})\,d{\bfm\Omega} \
,\label{radiationa}
\end{eqnarray}
where
\begin{eqnarray}
{\widehat J}_{_{i\ell}}({\bfm v},\,{\bfm\Omega})=f_{_\ell}({\bfm
v})\,\left[\alpha_{_{i\ell }}+\beta_{_{i\ell }}\,I_{_{i\ell
}}({\bfm\Omega})\right]-\beta_{_{i\ell }}\,f_{_i}({\bfm
v})\,I_{_{i\ell}}({\bfm \Omega}) \ .\label{rada}
\end{eqnarray}
By taking into account all the energy levels higher than $\ell$,
the loss term is due to absorption, while the gain term is due to
spontaneous and stimulated emission. The situation is reversed
when we consider all the energy levels lower than $\ell$.\\
The kinetic equation for photons $p_{ij}$ with intensity
$I_{_{ij}}$ reads \cite{Po}:
\begin{eqnarray}
\frac{\partial\,I_{_{ij}}}{\partial\,t}+{\bfm\Omega}\cdot{\bfm\nabla}I_{_{ij}}=
\omega_{_{ij}}\,{\widetilde J}_{_{ij}}({\bfm\Omega}) \ ,
\label{kinetica}
\end{eqnarray}
where
\begin{eqnarray}
{\widetilde J}_{_{ij}}({\bfm\Omega})=\int_{I\!\!R^3}{\widehat
J}_{_{ij}}({\bfm v},\,{\bfm\Omega})\,d{\bfm v} \ .
\end{eqnarray}
The gain term is due to spontaneous and stimulated
emission, while the loss term is due to absorption.\\

%%%%%%%%%%%%%%%%%%%%%%%%%%%%%%%%%%%%%%%%%%%%%%%%%%%%%%%%%%%%%%%%%%%%
\sect{}

We describe briefly two models used in the numerical simulations.
They take into account an inverse power law decay of the
distribution function with respect to energy
\cite{Abe,Kaniadakis0,Kaniadakis00}.\\

{\bf 1}) The first model that we consider has been proposed by
B\"uy\"ukkili\c c et al. \cite{5,6}. Relevant applications to the
blackbody problem can be found in \cite{W1,W2}. Preliminarily, we
define the $q$-deformed exponential and logarithm
\begin{eqnarray}
\exp_{_q}(x)=[1+(1-q)\,x]_{_+}^{1/(1-q)} \ ,\label{expq}\\
\ln_{_q}(x)={x^{1-q}-1\over 1-q} \ ,
\end{eqnarray}
where $[x]_+=x$ for $x\geq0$ and $[x]_+=0$ for $x<0$.  These
$q$-deformed functions have been introduced in thermo-statistics
for the first time in Ref. \cite{Tsallis}.\\
The characteristic functions $\Phi$ and $\Psi$ are given by
\cite{RSc}
\begin{eqnarray}
\Phi(x)=\exp\left[\ln_{_q}(x)\right] \ ,\\
\Psi(x)=\exp\left[\ln_{_q}(x)-\ln_{_q}\left({x\over 1+x}
\right)\right] \ ,
\end{eqnarray}
where $1\leq q<2$ and for $q=1$ the standard statistics is
recovered. For bosons we observe that the condition $\Psi(+\infty)=+\infty$ does not hold.\\
The deformed Bose-Einstein distribution function is given by
\begin{eqnarray}
I_{_{ij}}={\alpha_{_{ij}}\over\beta_{_{ij}}}\,\left[{1\over\exp_{_q}(-\omega_{_{ij}}/T)}-1\right]^{-1}
\ .
\end{eqnarray}

{\bf 2)} The second model that we consider has been proposed by
Kaniadakis \cite{11}. Preliminarily we define the
$\kappa$-deformed exponential and logarithm
\begin{eqnarray}
\exp_{_{\{\kappa\}}}(x)=\left(\sqrt{1+\kappa^2\,x^2}+\kappa\,x\right)^{1/\kappa}
 \ ,\\
\ln_{_{\{\kappa\}}}(x)={x^\kappa-x^{-\kappa}\over2\,\kappa} \ .
\end{eqnarray}
Remarkably the $\kappa$-exponential satisfy the property
$\exp_{_{\{\kappa\}}}(x)\,\exp_{_{\{\kappa\}}}(-x)=1$.The
characteristic functions $\Phi$ and $\Psi$ are given by \cite{RSc}
\begin{eqnarray}
\Phi(x)=\exp\left[\ln_{_{\{\kappa\}}}(x)\right] \ ,\\
\Psi(x)=\exp\left[\ln_{_{\{\kappa\}}}(x)-\ln_{_{\{\kappa\}}}
\left({x\over1+x}\right)\right] \ ,
\end{eqnarray}
where $0\le |k|<1$ and for $\kappa=0$ the standard statistics is
recovered. In the case of bosons, the condition for
$\Phi(+\infty)=+\infty$ is fulfilled as well. The deformed
Bose-Einstein distribution function is given by
\begin{eqnarray}
I_{_{ij}}={\alpha_{_{ij}}\over\beta_{_{ij}}}\,
\left[\exp_{_{\{\kappa\}}}\left(\frac{\omega_{_{ij}}}{T}\right)-1\right]^{-1}
\ .
\end{eqnarray}
%%%%%%%%%%%%%%%%%%%%%%%%%%%%%%%%%%%%%%%%%%%%%%%%%%%%%%%%%%%%%%%%%%%%%%%%%%%%%%%%%%%%%%

\vfill\eject
\end{document}